 \documentclass[usegraphicx,useAMS,referee]{biom}
\usepackage{amsmath}
\usepackage{amssymb}
\usepackage[svgnames]{xcolor}
\usepackage{float}
\usepackage{caption,subcaption}

\def\bSig\mathbf{\Sigma}

\newcommand{\commentout}[1]{}
\def\subopt{_{\scriptscriptstyle \sf opt}}
\def\subaDR{_{\scriptscriptstyle a, \sf DR}}
\def\suboneDR{_{\scriptscriptstyle 1, \sf DR}}
\def\subzeroDR{_{\scriptscriptstyle 0, \sf DR}}
\def\gopt{g\subopt}
\def\subgopt{_{\gopt}}
\def\subghat{_{\ghat}}
\def\subghatstar{_{\ghat^*}}
\def\fhat{\widehat{f}}
\def\PTE{\mbox{PTE}}
\def\PTEhat{\widehat{\PTE}}
\def\supone{^{(1)}}
\def\supzero{^{(0)}}
\def\supa{^{(a)}}
\def\Sbb{\mathbb{S}}
\def\Mbb{\mathbb{M}}
\def\bXvec{\vec{\bX}}

\def\bgamma{\boldsymbol{\gamma}}
\def\bgammahat{\widehat{\bgamma}}
\def\omegahat{\widehat{\omega}}
\def\Pschat{\widehat{\Psc}}
\def\Psc{\mathcal{P}}
\def\mhat{\widehat{m}}
\def\ghat{\widehat{g}}
\def\ghatstar{\widehat{g}^*}
\def\sumin{\sum_{i=1}^n}
\def\Deltahat{\widehat{\Delta}}
\def\muhat{\widehat{\mu}}
\def\pihat{\widehat{\pi}}
\def\psihat{\widehat{\psi}}
\def\bgamma{\boldsymbol{\gamma}}
\def\bgammahat{\widehat{\bgamma}}
\def\balpha{\boldsymbol{\alpha}}
\def\balphahat{\widehat{\balpha}}
\def\bbeta{\boldsymbol{\beta}}
\def\bbetahat{\widehat{\bbeta}}
\def\Hsc{\mathcal{H}}
\def\Fsc{\mathcal{F}}
\def\argmax{\mbox{argmax}}
\def\Dsc{\mathcal{D}}
\def\bXvec{\vec{\bX}}
\def\bxvec{\vec{\bx}}
\def\bPhi{\boldsymbol{\Phi}}
\def\Msc{\mathcal{M}}
\def\bzero{\mathbf{0}}
\def\bU{\mathbf{U}}
\def\bUhat{\widehat{\bU}}
\def\subam{_{a,\scriptscriptstyle \sf m}}
\def\subaf{_{a,\scriptscriptstyle \sf f}}
\def\lambdahat{\widehat{\lambda}}
\def\subDR{_{\scriptscriptstyle \sf DR}}

\def\subagDR{_{a,g,\scriptscriptstyle \sf DR}}

\def\subghatDR{_{\ghat\subDR,\scriptscriptstyle \sf DR}}
\def\subghatstarDR{_{\ghatstar\subDR,\scriptscriptstyle \sf DR}}

\def\suboneghatDR{_{1,\ghat\subDR,\scriptscriptstyle \sf DR}}
\def\subzeroghatDR{_{0,\ghat\subDR,\scriptscriptstyle \sf DR}}
\def\subaghatstarDR{_{a,\ghat\subDR^*,\scriptscriptstyle \sf DR}}
\def\suboneghatstarDR{_{1,\ghat\subDR^*,\scriptscriptstyle \sf DR}}
\def\subzeroghatstarDR{_{0,\ghat\subDR^*,\scriptscriptstyle \sf DR}}
\def\sumin{\sum_{i=1}^n}
\def\omegahatstar{\omegahat^*}
\def\ninv{n^{-1}}
\def\Msc{\mathcal{M}}
\def\Mschat{\widehat{\Msc}}

\def\Vscbar{\bar{\Vsc}}
\def\supb{^{\scriptscriptstyle [b]}}
\def\Vsc{\mathcal{V}}

\def\subFnaive{_{\scriptscriptstyle \sf F, naive}}
\def\subFbX{_{\scriptscriptstyle \sf F, \bX}}
\def\subFIPW{_{\scriptscriptstyle \sf F, IPW}}
\def\subP{_{\scriptscriptstyle \sf P}}
\def\subW{_{\scriptscriptstyle \sf W}}

\def\bmhat{\widehat{\mathbf{m}}}
\def\nnhalf{n^{-\half}}
\def\half{\frac{1}{2}}
\def\Dhat{\widehat{D}}
\def\bm{\mathbf{m}}
\def\bbf{\mathbf{f}}
\def\supdag{^{\scriptscriptstyle \dag}}

\def\Fbb{\mathbb{F}}
\def\bX{\mathbf{X}}
\def\bx{\mathbf{x}}
\def\trans{^{\scriptscriptstyle \sf T}}

\def\bBsc{\boldsymbol{\mathcal{B}}}
\def\Asc{\mathcal{A}}
\def\Gsc{\mathcal{G}}
\definecolor{jcolor}{RGB}{041,122,000}
\definecolor{darkred}{RGB}{100,000,000}
\definecolor{purple}{RGB}{200,000,200}

\def\subopt{_{\scriptscriptstyle \sf opt}}
\def\subaDR{_{\scriptscriptstyle a, \sf DR}}
\def\suboneDR{_{\scriptscriptstyle 1, \sf DR}}
\def\subzeroDR{_{\scriptscriptstyle 0, \sf DR}}
\def\gopt{g\subopt}
\def\subgopt{_{\gopt}}
\def\subghat{_{\ghat}}
\def\fhat{\widehat{f}}
\def\PTE{\mbox{PTE}}
\def\PTEhat{\widehat{\PTE}}
\def\supone{^{(1)}}
\def\supzero{^{(0)}}
\def\supa{^{(a)}}
\def\Sbb{\mathbb{S}}
\def\Mbb{\mathbb{M}}
\def\bXvec{\vec{\bX}}

\def\bgamma{\boldsymbol{\gamma}}
\def\bgammahat{\widehat{\bgamma}}
\def\omegahat{\widehat{\omega}}
\def\Pschat{\widehat{\Psc}}
\def\Psc{\mathcal{P}}
\def\mhat{\widehat{m}}
\def\ghat{\widehat{g}}
\def\sumin{\sum_{i=1}^n}
\def\Deltahat{\widehat{\Delta}}
\def\muhat{\widehat{\mu}}
\def\pihat{\widehat{\pi}}
\def\psihat{\widehat{\psi}}
\def\bgamma{\boldsymbol{\gamma}}
\def\bgammahat{\widehat{\bgamma}}
\def\balpha{\boldsymbol{\alpha}}
\def\balphahat{\widehat{\balpha}}
\def\bbeta{\boldsymbol{\beta}}
\def\bbetahat{\widehat{\bbeta}}
\def\Hsc{\mathcal{H}}
\def\Fsc{\mathcal{F}}
\def\argmax{\mbox{argmax}}
\def\Dsc{\mathcal{D}}
\def\bXvec{\vec{\bX}}
\def\bxvec{\vec{\bx}}
\def\bPhi{\boldsymbol{\Phi}}
\def\Msc{\mathcal{M}}
\def\bzero{\mathbf{0}}
\def\bU{\mathbf{U}}
\def\bUhat{\widehat{\bU}}
\def\subam{_{a,\scriptscriptstyle \sf m}}
\def\subaf{_{a,\scriptscriptstyle \sf f}}
\def\lambdahat{\widehat{\lambda}}
\def\subDR{_{\scriptscriptstyle \sf DR}}

\def\subghatDR{_{\ghat,\scriptscriptstyle \sf DR}}
\def\suboneghatDR{_{1,\ghat,\scriptscriptstyle \sf DR}}
\def\subzeroghatDR{_{0,\ghat,\scriptscriptstyle \sf DR}}
\def\bUsc{\boldsymbol{\Usc}}
\def\Usc{\mathcal{U}}
\def\ninv{n^{-1}}
\def\psibar{\bar{\psi}}
\def\pibar{\bar{\pi}}
\def\omegabar{\bar{\omega}}
\def\balphabar{\bar{\balpha}}
\def\subFnaive{_{\scriptscriptstyle \sf F, naive}}
\def\subFbX{_{\scriptscriptstyle \sf F, \bX}}
\def\subFIPW{_{\scriptscriptstyle \sf F, IPW}}
\def\subP{_{\scriptscriptstyle \sf P}}
\def\subW{_{\scriptscriptstyle \sf W}}

\def\bD{\mathbf{D}}
\def\eps{\varepsilon}

\def\Vscbar{\bar{\Vsc}}
\def\supb{^{\scriptscriptstyle [b]}}
\def\Vsc{\mathcal{V}}

\def\Msc{\mathcal{M}}
\def\Mschat{\widehat{\Msc}}

\title[Surrogates in Real World Data Settings]{On the Evaluation of Surrogate Markers in Real World Data Settings}

\author{Larry Han$^*$\email{larryhan@g.harvard.edu}, 
Xuan Wang$^{**}$\email{xwang@hsph.harvard.edu}, and 
Tianxi Cai$^{***}$\email{tcai@hsph.harvard.edu} \\
Department of Biostatistics, Harvard T.H. Chan School of Public Health, Boston, Massachusetts, 02115}

\begin{document}


\date{{\it Received April} 2021. {\it Revised August} 2021.  {\it
Accepted December} 2021.}



\pagerange{\pageref{firstpage}--\pageref{lastpage}} 
\volume{64}
\pubyear{2021}
\artmonth{April}


\doi{10.1111/j.1541-0420.2005.00454.x}


\label{firstpage}


\begin{abstract}
Shortcomings of randomized clinical trials are pronounced in urgent health crises, when rapid identification of effective treatments is critical. Leveraging short-term surrogates in real-world data (RWD) can guide policymakers evaluating new treatments. In this paper, we develop novel estimators for the proportion of treatment effect (PTE) on the true outcome explained by a surrogate in RWD settings. We propose inverse probability weighted and doubly robust (DR) estimators of an optimal transformation of the surrogate and PTE by semi-nonparametrically modeling the relationship between the true outcome and surrogate given baseline covariates. We show that our estimators are consistent and asymptotically normal, and the DR estimator is consistent when either the propensity score model or outcome regression model is correctly specified. We compare our proposed estimators to existing estimators and show a reduction in bias and gains in efficiency through simulations. We illustrate the utility of our method in obtaining an interpretable PTE by conducting a cross-trial comparison of two biologic therapies for ulcerative colitis.
\end{abstract}

%

\begin{keywords}
Double robustness; Proportion of treatment effect explained; Real world data; Semi-nonparametric estimation; Surrogate marker.
\end{keywords}


\maketitle


%

\section{Introduction}
\label{s:intro}

While randomized clinical trials (RCTs) remain the gold standard instrument for identifying efficacious and safe drugs (Concato, et al., 2000), RCTs typically require long-term follow-up of patients to observe a sufficient number of events to estimate treatment effects and even then, the outcomes may be costly to measure  (Bentley et al., 2019).  RCTs are also often limited to narrowly defined patient populations with results that are not always generalizable. These shortcomings are especially pronounced in urgent health crises and have led to increased interest in using real world data (RWD) and shorter term surrogate endpoints to efficiently and effectively evaluate treatments  (Hernán \&  Robins, 2016;  Corrigan-Curay et al.,  2018; Hey et al., 2020; Gyawali et al., 2020). For example, as multiple pharmaceutical companies pushed to develop vaccines and treatments for COVID-19 and to test them in RCTs  (Lurie et al., 2020), RWD from electronic health records (EHRs) have been collected at breakneck speed (Brat et al., 2020).

The use of valid surrogate markers to infer treatment effects on long term outcomes has the potential to reduce trial cost and study duration (Ciani et al., 2017; Wickström \& Moseley, 2017). The explosion in recent years of RWD highlights an untapped opportunity to identify and validate surrogate markers. Since Prentice (1989) originally proposed a definition and operational criteria for identifying valid surrogate markers, many statistical methods have been developed to make inference about the proportion of treatment effect (PTE)  explained by a surrogate in RCT settings  (Freedman \& Schatzkin, 1992; Lin et al., 1997; Wang \& Taylor, 2002; Parast et al., 2016; Price et al., 2018; Wang et al., 2020).
For example, Freedman \& Schatzkin (1992)  proposed a parametric model-based estimate assuming two regression models which rarely hold simultaneously (Lin et al., 1997). Wang \& Taylor (2002) proposed alternative measures of PTE that examined what the treatment effect would have been if the surrogate had the same distribution across treatment groups. 
Parast et al. (2016) proposed a fully nonparametric estimation procedure for the PTE defined in Wang \& Taylor (2002). More recently, Wang et al. (2020) proposed an alternative non-parametric PTE estimator by identifying an optimal transformation of the surrogate $S$, $\gopt: S \to \gopt(S)$, such that $\gopt(S)$ optimally predicts $Y$. In addition to requiring weaker assumptions than those required by Parast et al. (2016), this approach has the advantage of providing a direct approximation to the treatment effect on the outcome $Y$ using the treatment effect on  $\gopt(S)$. 

These existing PTE estimates are derived for data from RCTs and not directly applicable to RWD where treatment assignment $A$ may depend on confounding factors $\bX$. In this paper, we follow the strategy of Wang et al. (2020) to define PTE based on $\gopt$ and propose both IPW and doubly robust PTE estimators using RWD  by semi-non-parametrically modeling the relationship between $Y$ and $S$ given $\bX$ and imposing a propensity score (PS) model for $A \mbox{ given } \bX$. We propose perturbation resampling methods for variance and confidence interval estimation. We establish the asymptotic properties of the proposed estimators, including double robustness of the proposed estimator in that it is consistent when either the PS model or the outcome regression (OR) models is correctly specified. Our simulation studies demonstrate that the proposed estimators and inference procedures perform well in finite samples. Finally, we illustrate the utility of our proposed procedures by conducting a cross-trial comparison of two biologic therapies for ulcerative colitis (UC).

\section{Methods}

\subsection{Setting and notations}

Let $Y$ be the primary outcome and $S$ be the surrogate marker, both of which may be discrete or continuous. Throughout, the notation takes $S$ to be continuous, but all derivations and theoretical results remain valid if $S$ is discrete by replacing density functions with probability mass functions. We denote $\{Y^{(a)}, S^{(a)}\}$ as the respective potential primary outcome and surrogate marker under treatment $A=a$, where $A = 1$ and $A=0$ denote the treatment and the control group, respectively. With RWD, only $Y_i = A_iY_i^{(1)} + (1-A_i)Y_i^{(0)}$ and $S_i = A_i S_i^{(1)} + (1-A_i)S_i^{(0)}$ can be observed for an individual $i$, and the treatment assignment $A_i$ may depend on baseline confounding factors $\bX_i$.  For identifiability, we require the standard assumptions (Rubin, 2005; Imbens \& Rubin, 2015):
\begin{equation}
{\pi_a(\bx) \equiv P(A=a|\bX=\bx) \in (0,1)}
\label{eq:ps}
\end{equation}
\begin{equation}
\left(Y^{(1)}, Y^{(0)}, S^{(1)}, S^{(0)}\right) \perp A \mid \bX
\label{eq:nuc}
\end{equation}
Assumption~(\ref{eq:ps}) states that within all covariate levels, patients may receive either treatment so that the PS is bounded away from $0$ and $1$. Assumption~(\ref{eq:nuc}) implies that $\bX$ includes all confounders that can affect the primary outcome and treatment simultaneously, or the surrogate and treatment simultaneously (Rubin, 2005; Imbens \& Rubin, 2015). We assume that the RWD for analysis consist of $n$ independent and identically distributed random variables $\{\mathbf{D_i} = (Y_i, S_i, A_i, \bX_i\trans)\trans, i=1,...,n\}$.

\subsection{Target parameter and leveraging surrogates}

The average treatment effect on $Y$ is defined as:
$$
\Delta=\mu_{1}-\mu_{0}, \quad \text { where } \mu_{a}=E(Y^{(a)}) = \int E(Y \mid A = a, \bX)d \Fbb(\bX),$$
and $\Fbb(\bx) = P(\bX\le \bx)$. Without loss of generality, we assume that $\Delta > 0$.  To approximate $\Delta$ based on the treatment effect on $S$,  Wang et al. (2020) identified a transformation function $\gopt(\cdot)$ such that the treatment effect on the transformed surrogate, $\Delta\subgopt = E[g_{opt}(S^{(1)})- g_{opt}(S^{(0)})],$ can optimally predict $\Delta$ in a certain sense. More formally, the optimality of $\gopt$ is with respect to minimizing the mean squared error
\begin{align*}
\mathcal{L}_{oracle}(g_{opt}) =
E\left[\left(Y^{(1)}-Y^{(0)}\right)-\left\{g_{{\text {opt}}}\left(S^{(1)}\right)-g_{\text {opt}}\left(S^{(0)}\right)\right\}\right]^{2} 
\end{align*}
under the {\em working assumption} of $(Y^{(1)}, S^{(1)}) \perp (Y^{(0)}, S^{(0)})$. It was shown that $g_{opt}$ takes the form 
$$g_{opt}(s) = m(s) + \lambda \mathcal{P}_0(s) \quad \mbox{with}\quad
m(s) = m_1(s) \mathcal{P}_1(s) + m_0(s) \mathcal{P}_0(s),
$$
where $m_a(s) = E(Y^{(a)}|S^{(a)}=s)$, $\mathcal{P}_a(s) = {f_a(s)}(f_0(s) + f_1(s))^{-1}$, $f_a(s) = {dF_a(s)}/{ds},$ $$ \ \lambda=\frac{\int \{m_0(s)-m_1(s)\} \mathcal{P}_1(s) dF_0(s)}{\int \mathcal{P}_0(s) dF_0(s)} = \frac{\mu_0 - \int m(s) dF_0(s)}{\int \mathcal{P}_0(s) dF_0(s)},
$$ 
and $F_a(s) = P(S^{(a)}\leq s)$. In addition, by employing the transformation $\gopt$ and defining the PTE of $S$ as
$\PTE\subgopt \equiv {\Delta\subgopt}/{\Delta} , $
Wang et al. (2020) showed that even if the working independence assumption does not hold, $\PTE\subgopt  \in [0,1]$ provided that
\begin{align*}
    & \text{(A1)} \quad \Sbb_1(u) \geq \Sbb_0(u) \quad \mbox{for all $u$}, \\
& \text{(A2)} \quad \Mbb_1(u) \geq \Mbb_0(u) \quad \mbox{for all $u$ in the common support of $g_{opt}(S^{(1)})$ and  $g_{opt}(S^{(0)})$},
\end{align*}

where $\Sbb_a(u) = P\{g_{opt}(S\supa) \geq u \}$ and $\Mbb_a(u) = E(Y^{(a)}|g_{opt}(S^{(a)}) = u)$, for $a=0,1$. 
Assumptions (A1) and (A2) are  weaker than those required in the literature to ensure that the PTE is between 0 and 1 and hence to avoid the surrogate paradox (VanderWeele, 2013). Our goal here is to construct robust estimates for $\gopt$ and $\PTE\subgopt$ using RWD. 

\section{Two Proposed Estimation Methods}
Estimation of $g_{opt}(s)$ and $\PTE\subgopt$ using RWD is more challenging than using RCT data because we cannot directly estimate $m(s)$, $\lambda$, and $\mathcal{P}_a(s)$ due to confounding. We propose an inverse probability weighted (IPW) estimator and a doubly robust (DR) estimator for $\gopt$ and $\PTE\subgopt$ accounting for the effects of $\bX$ on $A$, $Y$ and $S$. We first present the simpler IPW estimator and then the DR estimator. For both estimators, we impose a parametric model for $\pi_1(\bX)$, denoted by $\pi_1(\bX; \balpha)$, where $\balpha$ is a finite dimensional parameter that can be estimated as the standard maximum likelihood estimator, $\balphahat$. A simple example is a logistic regression model $\pi_1(\bX; \balpha) = G\{\balpha\trans\bPhi(\bX)\}$, where $G(x)=e^x/(1+e^x)$ and $\bPhi(\bX)$ is a vector of basis functions of $\bX$ to  account for potential non-linear effects. 

\subsection{IPW Estimation}
To construct an IPW estimator for $\gopt$, we first obtain IPW kernel smoothed estimators for $m_a(s)$ and $f_a(s)$  respectively as
$$\mhat_a(s) = \frac{\sum_{i=1}^n K_h(S_i-s)Y_i \omegahat_{ai}}{\sum_{i=1}^n K_h(S_i-s)\omegahat_{ai}} \quad \mbox{and}\quad \fhat_a(s) = \frac{\sum_{i=1}^n K_h(S_i-s)\omegahat_{ai}}{\sum_{i=1}^n \omegahat_{ai}},$$
where $\omegahat_{ai} = I(A_i=a) / \pi_a(\bX_i,\balphahat),$ $K_h(\cdot)=h^{-1}K(\cdot/h)$, $K(\cdot)$ is a symmetric density function and $h=O(n^{-\nu})$ with $\nu \in (1/4, 1/2)$. Then $m(\cdot)$, $
\Psc_a(\cdot)$ and $\lambda$ may be estimated as
$$\mhat(s) = \sum_{a=0}^1 \mhat_a(s) \Pschat_a(s) , \quad \Pschat_a(s) = \frac{\fhat_a(s)}{\fhat_1(s) + \fhat_0(s)},\quad \hat{\lambda} = \frac{\int (\mhat_0(s)-\mhat_1(s))\Pschat_1(s) \fhat_0(s) ds}{\int \Pschat_0(s)\fhat_0(s)ds}, $$
respectively. Subsequently, we construct plug-in estimators for $\gopt(s)$, $\Delta\subgopt$, $\Delta$ and $\PTE\subgopt$ as 
$$
\ghat(s) = \mhat(s) + \hat{\lambda} \Pschat_0(s), \quad 
\widehat{\Delta}\subghat = \muhat_{1,\ghat} - \muhat_{0,\ghat}, \quad
\Deltahat = \muhat_1 - \muhat_0,\quad \mbox{and}\quad 
\PTEhat_{\ghat} = \frac{\widehat{\Delta}\subghat}{\widehat{\Delta}},
$$ 
where $$\muhat_{a,g} =  \frac{\sumin g(S_i)\omegahat_{ai}}{\sumin \omegahat_{ai}}
 \quad \mbox{and}\quad\muhat_a =  \frac{\sum_{i=1}^n Y_i\omegahat_{ai}}{\sumin \omegahat_{ai}.}$$

We show in Appendix 1 of the supplementary materials that when $\pi_1(\bx; \balpha)$ is correctly specified, $\PTEhat_{\ghat}$ is consistent for  $\PTE\subgopt$. We also show that $\sqrt{n}(\PTEhat\subghat-\PTE\subgopt)$ converges in distribution to a normal distribution with mean $0$ and variance $\sigma^2$, where the form of $\sigma^2$ is given in the supplementary materials. 

\subsection{Doubly Robust Estimation}
When the PS model is incorrectly specified, the IPW estimator is likely to be biased. Here, we propose augmented IPW estimators for $\gopt$ and $\PTE\subgopt$ to achieve improved robustness and efficiency. Following Robins et al. (1994), for any counterfactual random variable $U^{(a)}$, an augmented IPW estimator for its mean $E(U\supa)$ can be constructed as
$$ n^{-1} \sum_{i=1}^n \left\{ \omegahat_{ai}U_i - (\omegahat_{ai}-1)  \widehat{\phi}_a(\bX_i) \right\}, $$
where $\widehat{\phi}_a(\bX_i)$ is an estimator for $E(U_i\supa \mid \bX_i)$ derived under a specified model. This estimator is doubly robust in the sense that it is consistent for $E(U\supa)$ when either the PS model for $\pi_a(\bX)$ or the outcome model for $E(U_i\supa \mid \bX_i)$ is correctly specified. Deriving an augmented IPW estimator for $\PTE\subgopt$ is more involved since $\gopt(S)$ involves conditional mean functions of $Y\supa \mid S\supa$ and density functions of $S\supa$ for $a=0,1$. 

 To construct a DR estimator for $\gopt(s) = m(s) + \lambda \mathcal{P}_0(s)$, we propose the following DR estimators for 
$m_a(s)$ and $f_a(s)$ respectively,
\begin{subequations}
\begin{align}
    \mhat\subaDR(s) &= \frac{\Mschat\subaDR(s)}{\fhat\subaDR(s)} \label{eq-mhatDR}, \\
    \Mschat\subaDR(s) &= n^{-1} \sum_{i=1}^n \left\{K_h(S_i-s) Y_i \omegahat_{ai}- (\omegahat_{ai}-1) \psihat\supdag\subam(s; \bX_i)\psihat\supdag\subaf(s; \bX_i) \right\} \label{eq-MshatDR}, \\
    \fhat\subaDR(s)  &= n^{-1} \sum_{i=1}^n \left\{K_h(S_i-s)\omegahat_{ai} - (\omegahat_{ai}-1) \psihat\subaf(s; \bX_i)  \right\},  \label{eq-fhatDR}
\end{align}
\end{subequations}

where 
$\psihat\subam(\bx)$ and $\psihat\subaf(s; \bx)$ are the respective estimators for 
\begin{align*}
\psi\subam(s; \bx) & = E(Y_i\supa \mid S_i\supa = s, \bX_i=\bx) = E(Y_i \mid A_i=a, S_i=s, \bX_i = \bx) \ \mbox{and } \\
\psi\subaf(s; \bx) & = \frac{\partial P(S_i\supa \le s \mid \bX_i = \bx)}{\partial s} ,
\end{align*}
In Appendix 2 of the supplementary materials, we show that $\mhat\subaDR(s)$ and $\fhat\subaDR(s)$ are consistent for $m_a(s)$ and $f_a(s)$ if either $\sup_{\bx}|\pihat_a(\bx)-\pi_a(\bx)|\to 0$ in probability or $\sup_{\bx,s}\{|\psihat\subam(s; \bx)-\psi\subam(s; \bx)|+|\psihat\subaf(s; \bx)-\psi\subaf(s; \bx)|\}\to 0$ in probability. 

To construct estimators $\psihat\subam(s;\bx)$ and $\psihat\subaf(s; \bx)$, we impose flexible semi-non-parametric models for $Y\supa \mid S\supa, \bX$ and $S\supa \mid \bX$ to minimize assumptions on the dependency structure between $S$ and $Y$. Specifically,  we first impose a generalized regression model (GRM) (Han, 1987) for $S_i \mid A_i=a, \bX_i$:
$$S_i  = \Dsc_a \odot \Hsc_a (\bX_i\trans\bgamma_{a} , \epsilon_{ia})  \quad\mbox{with} \quad P(\epsilon_{ai} \le e \mid \bX_i) = \Fsc_a(e) ,$$
where $\Dsc_a(\cdot)$ is an increasing function and $\Hsc_a(\cdot,\cdot)$ is a strictly increasing function of each of its arguments, and the unknown covariate effects 
$\bgamma_a = (\gamma_{a1}, ...,\gamma_{ap}) \trans$ are constrained to the unit sphere $\Omega: \{\bgamma: \|\bgamma\|_2 = 1\}$ for identifiability. With the given $\bgamma_a$ under GRM and the no-unmeasured-confounders assumption,  $\psi\subaf(s;\bx)$ can be estimated non-parametrically via kernel smoothing. To estimate $\bgamma_a$, Sherman (1993) showed that the maximum rank correlation estimator 
$$\bgammahat_a = \argmax_{\bgamma  \in \Omega} \left\{\sum_{i \neq j, A_i=A_j=a} I(\bX_i \trans \bgamma > \bX_j \trans \bgamma) I(S_i > S_j)\right\}$$ 
is consistent and asymptotically normal for $\bgamma_a$. Subsequently, we estimate $\psi\subaf(s,\bx)$ as
\begin{equation}
   \psihat\subaf(s; \bx) = \frac{\sumin K_\zeta(\bgammahat_a\bX_i - \bgammahat_a\trans\bx)K_h(S_i-s)}{\sumin K_\zeta(\bgammahat_a\bX_i - \bgammahat_a\trans\bx)} . 
\end{equation}

To estimate $\psi\subam(s; \bx)$, we impose a varying-coefficient generalized linear model (VGLM) (Hastie \& Tibshirani, 1993):
$$E(Y_i\ \mid A_i =a, S_i = s, \bX_i) = M\{\bbeta_a(S_i)\trans\bXvec_i \} ,$$ 
where $M(\cdot)$ is a known smooth link function, $\bxvec = (1,\bx\trans)\trans$ for any vector $\bx$ and $\bbeta_a(s)$ is an unknown $p+1$ dimensional unspecified smooth coefficient functions. We may estimate $\bbeta_a(s)$ as $\bbetahat_a(s)$, the solution to
$$
\bUhat_a(\bbeta; s) = \ninv \sumin I(A_i = a)K_h(S_i-s) \bXvec_i\left\{Y_i - M(\bbeta\trans\bXvec_i)\right\} = \bzero
$$
Then we estimate $\psi\subam(s; \bx)$ as
\begin{equation}
    \psihat\subam(s, \bx) = M\{\bbetahat_a(s)\trans\bxvec\}.
\end{equation}

These estimators (4) and (5) can then be plugged into (\ref{eq-MshatDR}) and (\ref{eq-fhatDR}) to construct $\fhat\subaDR(s)$ and $\mhat\subaDR(s)$ as in (\ref{eq-mhatDR}). 

Based on $\mhat\subaDR(s)$ and $\fhat\subaDR(s)$, we obtain a doubly robust estimator for $g_{opt}(s)$ as:
$$\ghat\subDR(s) = \mhat\subDR(s) + \lambdahat\subDR\Pschat\subzeroDR(s),$$
where $\mhat\subDR(s) = \sum_{a=0}^1 \mhat\subaDR(s)\Pschat\subaDR(s)$, 
$$\lambdahat\subDR = \frac{\int\left\{\mhat\subzeroDR(s)-\mhat\suboneDR(s)\right\} \Pschat\suboneDR(s)  \fhat\subzeroDR(s)ds}{\int \Pschat\subzeroDR(s)  \fhat\subzeroDR(s) ds}, \quad \mbox{and} \quad \Pschat\subaDR(s) = \frac{\fhat\subaDR(s)}{\fhat\subzeroDR(s) + \fhat\suboneDR(s)}, 
$$
for $a=0,1$. We can now construct a plug-in estimator for $\Delta_{\gopt}$ as $\widehat{\Delta}\subghatDR= \muhat\suboneghatDR - \muhat\subzeroghatDR,$
where  
$$\muhat\subagDR = n^{-1}\sumin \left\{ g(S_i)\omegahat_{ai} - (\omegahat_{ai}-1)\hat{\zeta}_{a,g}(\bX_i) \right\} ,
$$
and $\hat{\zeta}_{a,g}(\bx) = \int g(s) \psihat\subaf(s,\bx)ds$ is an estimator for $\zeta_{a,g}(\bx) = E\{g(S_i^{(a)}) \mid \bX_i = \bx\}$ derived under the GRM. Similarly, we obtain $\widehat{\Delta}\subDR = \muhat\suboneDR - \muhat\subzeroDR$ to estimate $\Delta$, where 
$$\muhat\subaDR = n^{-1} \sumin \left\{ Y_i\omegahat_{ai} - (\omegahat_{ai}-1)\hat{\zeta}_a(\bX_i) \right\},
$$ 
where $\hat{\zeta}_a(\bx)=\int \psihat\subam(s;\bx)\psihat\subaf(s;\bx)ds$ is an estimator for $\zeta_a(\bx) = E(Y_i^{(a)} \mid \bX_i = \bx)$. Finally, we estimate $\PTE\subgopt$ as $\PTEhat\subghatDR = {\widehat{\Delta}\subghatDR} / {\widehat{\Delta}}\subDR.$ Following similar arguments as given in Appendix 2 of the supplementary materials, it is not difficult to show that $\Deltahat\subghatDR$, $\Deltahat\subDR$ and $\PTEhat\subghatDR$ are doubly robust estimators for $\Delta\subgopt$, $\Delta$, and $\PTE\subgopt$, respectively.

\section{Perturbation Resampling}

We propose to estimate the variability and construct confidence intervals of our proposed estimators using a perturbation-resampling approach (Jin et  al., 2001; Tian et al., 2005). For resampling, we generate $\{\mathbf{V}\supb = (V_1\supb,...,V_n\supb)\trans, b=1,...,B\}$, which are $n \times B$ independent and identically distributed non-negative random variables from a known distribution with unit mean and unit variance, such as the unit exponential distribution. For the IPW estimators, for each set of $\mathbf{V}=(V_1, ..., V_n)\trans$, we let $\Vscbar_i=V_i/(\ninv\sumin V_i)$,
$$\mhat_a^*(s) = \frac{\sum_{i=1}^n K_h(S_i-s)Y_i\Vscbar_i  \omegahat_{ai}^*}{\sum_{i=1}^n K_h(S_i-s)\Vscbar_i\omegahat_{ai}^*}  ,  \quad  \fhat_a^*(s) = \frac{\sum_{i=1}^n K_h(S_i-s)\Vscbar_i\omegahat_{ai}^*}{\sum_{i=1}^n\Vscbar_i \omegahat_{ai}^*} , \quad \omegahat_{ai}^* = \frac{I(A_i=a)}{\pi(\bX_i,\balphahat^*)},$$
where $\balphahat^*$ is obtained by fitting a weighted logistic regression $A_i \sim G\{\balpha\trans\bPhi(\bX_i)\}$ with weights $\{\Vscbar_i\}$. The perturbed counterparts of $\mhat(\cdot)$, $
\hat{\Psc}_a(\cdot)$ and $\widehat{\lambda}$ are obtained as
$$\mhat^*(s) = \sum_{a=0}^1 \mhat_a^*(s) \Pschat_a^*(s) , \quad \Pschat_a^*(s) = \frac{\fhat_a^*(s)}{\fhat_1^*(s) + \fhat_0^*(s)},\quad \widehat{\lambda}^* = \frac{\int \{\mhat_0^*(s)-\mhat_1^*(s)\}\Pschat_1^*(s) \fhat_0^*(s) ds}{\int \Pschat_0^*(s)\fhat_0^*(s)ds}, $$
respectively. Subsequently, we construct the perturbed counterparts of $\ghat(s)$, $\widehat{\Delta}\subgopt$, $\widehat{\Delta}$ and $\widehat{\PTE}$ as 
$$
\ghatstar(s) = \mhat^*(s) + \hat{\lambda}^* \Pschat_0^*(s) \quad 
\widehat{\Delta}^*\subghatstar = \muhat^*_{1,\ghatstar} - \muhat^*_{0,\ghatstar}, \quad
\Deltahat^* = \muhat^*_1 - \muhat^*_0,\quad \mbox{and}\quad 
\PTEhat^*_{\ghatstar} = \frac{\widehat{\Delta}^*\subghatstar}{\widehat{\Delta}^*},
$$ 
where 
$$\muhat^*_{a,g} = \frac{\sumin g(S_i)\Vscbar_i\omegahat^*_{ai}}{\sumin\Vscbar_i\omegahat^*_{ai}} \quad \mbox{and}\quad\muhat^*_a = \frac{\sum_{i=1}^n Y_i\Vscbar_i \omegahat_{ai}^*}{\sum_{i=1}^n \Vscbar_i \omegahat_{ai}^*}.$$

For the DR estimators, for each set of $\mathbf{V}$, we let
\begin{align*}
\mhat^*\subaDR(s) &= \frac{\Mschat^*\subaDR(s)}{\fhat^*\subaDR(s)},\\
\Mschat^*\subaDR(s) &= n^{-1} \sum_{i=1}^n  { \Vscbar_i} \left\{K_h(S_i-s) Y_i \omegahatstar_{ai}- (\omegahatstar_{ai}-1) \psihat^*\subam(s; \bX_i)\psihat^*\subaf(s; \bX_i) \right\}, \\
\fhat^*\subaDR(s) = & \ninv \sum_{i=1}^n { \Vscbar_i} \left\{K_h(S_i-s)\omegahatstar_{ai} - (\omegahatstar_{ai}-1) { \hat{\psi}_{a,f}^*(\bX_i)}  \right\}, 
\end{align*}
where  
$$
\psihat^*\subaf(s, \bx) = \frac{\sumin \Vscbar_i K_\zeta(\bX_i\trans\bgammahat^*_a - \bx\trans\bgammahat_a^*)K_h(S_i-s)}{\sumin \Vscbar_i K_\zeta(\bgammahat^*_a\bX_i - \bx\trans\bgammahat_a^*)} , \quad
\psihat^*\subam(s, \bx) = M\{\bbetahat_a^*(s)\trans\bxvec\}
$$ 
$\bgammahat_a^* = \argmax_{\bgamma  \in \Omega} \{\sum_{i \neq j, A_i=A_j=a} \Vscbar_i\Vscbar_j I(\bX_i \trans \bgamma > \bX_j \trans \bgamma) I(S_i > S_j)\}$ and
$\bbetahat_a^*(s)$ is the solution to
$$
\bUhat_a^*(\bbeta; s) \equiv \ninv \sumin \Vscbar_i I(A_i = a)K_h(S_i-s) \bXvec_i\left\{Y_i - M(\bbeta\trans\bXvec_i)\right\} = \bzero .
$$
We construct the perturbed counterparts of $\ghat\subDR(s)$, $\hat{\Delta}\subDR$, $\hat{\Delta}\subghatDR$, and $\PTEhat\subghatDR$ respectively as: 
$$\ghat\subDR^*(s) = \mhat\subDR^*(s) + \lambdahat\subDR^*\Pschat\subzeroDR^*(s),\quad
\Deltahat\subDR^* = \muhat\suboneDR^* - \muhat\subzeroDR^*,\quad
\Deltahat\subghatstarDR^* = \muhat\suboneghatstarDR^* - \muhat\subzeroghatstarDR^*, 
$$
and $\PTEhat\subghatstarDR^* = \widehat{\Delta}\subghatstarDR^*/\widehat{\Delta}\subDR^*$, where
$\mhat\subDR^*(s)=\sum_{a=0}^1 \mhat\subaDR^*(s)\Pschat^*\subaDR(s)$,
$$\lambdahat\subDR^* = \frac{\int\left\{\mhat\subzeroDR^*(s)-\mhat^*\suboneDR(s)\right\} \Pschat\suboneDR^*(s)  \fhat^*\subzeroDR(s)ds}{\int \Pschat^*\subzeroDR(s)  \fhat^*\subzeroDR(s) ds}, \quad  \Pschat^*\subaDR(s) = \frac{\fhat^*\subaDR(s)}{\fhat^*\subzeroDR(s) + \fhat^*\suboneDR(s)}, 
$$
$$
\muhat\subaghatstarDR^* = n^{-1}\sumin \Vscbar_i\left\{ \ghatstar\subDR(S_i)\omegahat_{ai}^* - (\omegahat_{ai}^*-1)\hat{\zeta}_{a,\ghatstar\subDR}^*(\bX_i) \right\},$$
$$
\muhat\subaDR^* = n^{-1} \sumin \Vscbar_i \left\{ Y_i\omegahat_{ai}^* - (\omegahat_{ai}^*-1)\hat{\zeta}_a^*(\bX_i) \right\},
$$
$\hat{\zeta}_{a,g}^*(\bx) = \int g(s) \psihat\subaf^*(s,\bx)ds$, and  $\hat{\zeta}_a^*(\bx)=\int \psihat^*\subam(s;\bx)\psihat^*\subaf(s;\bx)ds$.

Operationally, we generate a large number, say $B = 500$, realizations for $\mathbf{V}$ and then obtain $B$ realizations of the perturbed statistics of interest. Standard error estimates and confidence intervals can then be constructed based on empirical variances of these realizations.

\section{Simulation Studies}

We have conducted simulation studies to evaluate the finite sample performance of our proposed estimators compared to several existing methods. Namely, we considered the naive PTE estimator of Freedman et al. (1992), denoted $\PTEhat\subFnaive$, which does not take into account baseline covariates $\bX$ in the models $Y \mid S,A$ and $Y \mid A$; a modified version that incorporates $\bX$ into both models $Y \mid S, A, \bX$ and $Y \mid A, \bX$, denoted $\PTEhat\subFbX$; a modified version that incorporates the propensity score into both models, denoted $\PTEhat\subFIPW$; the PTE estimator given in Parast et al. (2016), denoted $\PTEhat\subP$; and the PTE estimator of Wang et al. (2020), denoted $\PTEhat\subW$. We let $n=400$ and $1000$ and choose $K(\cdot)$ as a Gaussian kernel. To obtain the bandwidth $h$ that satisfies the undersmoothing assumption, we set $h = h_{opt}n^{-c_0}$, $c_0 = 0.11$ where $h_{opt} = 1.06n^{-1/5}$ as in Scott (2015). We compute the true population parameters via Monte Carlo, under the counterfactual models used to generate the data, with $N=100,000$ averaged over $100$ replications. All results are summarized based on $500$ simulated datasets for each configuration, and $B=500$ resampling replications were used for variance and interval estimation based on the empirical variances.

We consider two general settings with a moderately strong surrogate in the first setting and a weak surrogate in the second setting. Specifically, in the first setting, 
we generate a 3-dimensional baseline covariate vector $\bX_i = (X_{i1}, X_{i2}, X_{i3})\trans$ as $X_{i1} \sim \mathcal{N}(0,0.04)$, $X_{i2} \sim \mbox{Gamma}(2,2)$ and $X_{i3} \sim \mbox{Uniform}(-1,1)$, and
\begin{align*}
    S_i^{(0)} &= \boldsymbol{\gamma}_{0[1]} \trans \bXvec_i + {\epsilon_i}, \quad S_i^{(1)} = \boldsymbol{\gamma}_{1[1]} \trans \bXvec_i + {\epsilon_i}, \\
    Y_i^{(0)} &= 0.5 S_i^{(0)} + \boldsymbol{\beta}_{0[1]} \trans \bXvec_i + X_{1i} X_{2i} + X_{2i} X_{3i} + {e_i}, \\
    Y_i^{(1)} &= 0.3 S_i^{(1)} + \boldsymbol{\beta}_{1[1]} \trans \bXvec_i + X_{1i} X_{2i} + X_{2i} X_{3i} + {e_i},
\end{align*}
where $\epsilon_i \sim \mathcal{N}(0,1)$, $e_i \sim \mathcal{N}(0,0.04)$, $\boldsymbol{\gamma}_{0[1]} = (0, 0.5, 1, -0.5) \trans$, $\boldsymbol{\gamma}_{1[1]} = (0, 1,0.5,2) \trans$, $\boldsymbol{\beta}_{0[1]} = (0, 0.2,-0.3,-0.5) \trans$, and $\boldsymbol{\beta}_{1[1]} = (0, 1,-0.5,0.2) \trans$. We generate $A_i \mid \bX_i$ from the propensity score model 
\begin{align}
    P(A_i=1 \mid \bX_i) = \mbox{expit}\{-0.8X_{i1} + 0.7X_{i2} - \log(X_{i3}) + 0.6X_{i1}X_{i3}\}. \label{sim-PS}
\end{align}
 Under this setting, $\Delta\subgopt = 0.29$ and $\Delta = 0.54$ so that the true potential outcomes PTE is $0.537$, i.e., $S$ is a moderately strong surrogate for $Y$. We consider scenarios in which we correctly specify both the PS and OR models, misspecify the PS model by omitting the interaction term $X_2 X_3$, misspecify the OR model by omitting the variable $X_2$ and all interaction terms including $X_2$, and misspecify both models.

In the second setting, we consider a relatively weak surrogate generate data such that the effect of $S$ on $Y$ is non-linear. We generate baseline covariates $\bX_i = (X_{i1},X_{i2},X_{i3})\trans$ from $X_{i1} \sim \mathcal{N}(0,1)$, $X_{i2} \sim \mbox{Gamma}(2,2)$, and $X_{i3} \sim \mbox{Uniform}(0,5)$. Given $\bX_i$, we generate $S$ and $Y$ from
\[
\begin{aligned}
    S_i^{(0)} =  \boldsymbol{\gamma}_{0[2]} \trans \bXvec_i + {\epsilon},& \quad S_i^{(1)} =  \boldsymbol{\gamma}_{1[2]} \trans \bXvec_i + {\epsilon_i}, \\
    Y_i^{(0)} = 100 + \boldsymbol{\beta}_{0[2]}(S_i^{(0)}) \trans \bX_i + e_i,& \quad Y_i^{(1)} = 50 + \boldsymbol{\beta}_{1[2]}(S_i^{(1)}) \trans \bX_i + e_i,
\end{aligned}
\]
where $\epsilon_i \sim \mathcal{N}(0,4)$ and $e_i \sim \mathcal{N}(0,1)$ and we let $\boldsymbol{\gamma}_{0[2]} = (100, 1, 5, 0) \trans$, $\boldsymbol{\gamma}_{1[2]} = (100,2,4,0) \trans$, $\boldsymbol{\beta}_{0[2]}(s) = (s,-2\log(s),25) \trans$, and $\boldsymbol{\beta}_{1[2]}(s) = (s, - 3\log(s), -14) \trans$. We generate $A_i \mid\bX_i$ from model (\ref{sim-PS}) as in setting 1. Under this data generating mechanism, $\Delta\subgopt = 5.7$ and $\Delta = 26.7$, resulting in $\PTE\subgopt = 0.214$. We consider scenarios in which we correctly specify both the PS and OR models, misspecify the PS model by omitting the $\log(X_3)$ term, misspecify the OR model by omitting $X_2$, and misspecify both models.

We first summarize results for setting 1. In Figure 1, we plot the empirical biases, the empirical standard error (ESE) compared to the average of the estimated standard error (ASE), and empirical coverage probabilities of the 95\% pointwise confidence intervals (CIs) for $g_{opt}(\cdot)$ based on the DR estimator $\ghat\subDR(\cdot)$ estimated with sample size $n=1000$ when (A) both the PS model and OR model are correctly specified, (B) the PS model is misspecified but the OR model is correctly specified, (C) the OR model is misspecified but the PS model is correctly specified, and (D) both models are misspecified. When at least one of the two models is correctly specified, the point estimates for
$g_{opt}(\cdot)$ present negligible bias, the ASEs are close to the ESEs, probabilities of the $95\%$ CIs are close to their nominal level. When both models are misspecified, bias is observed in the tails, the ASE somewhat underestimates the ESE, and the coverage probabilities of the $95\%$ confidence intervals are somewhat below the nominal level. Results for $n = 400$ bear similar patterns  and are hence omitted for brevity.

\begin{figure}
    \begin{center}

  \includegraphics[scale=0.5]{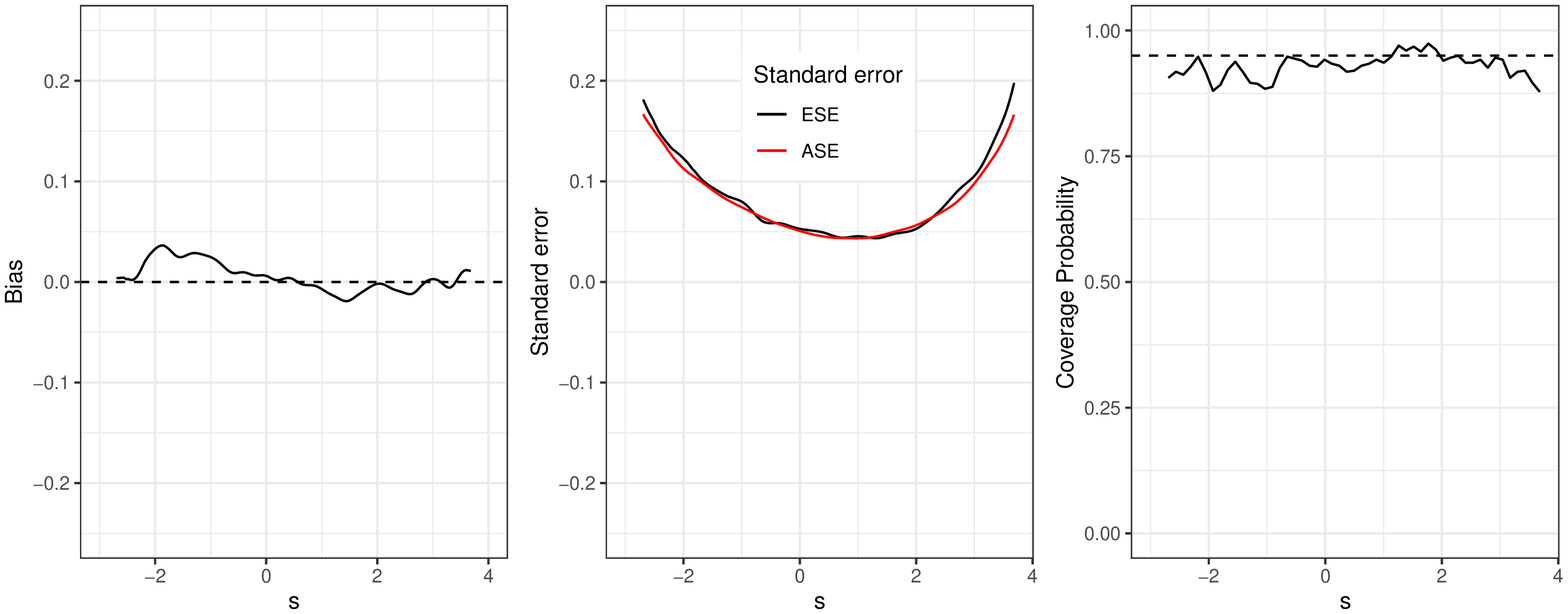}

  \includegraphics[scale=0.5]{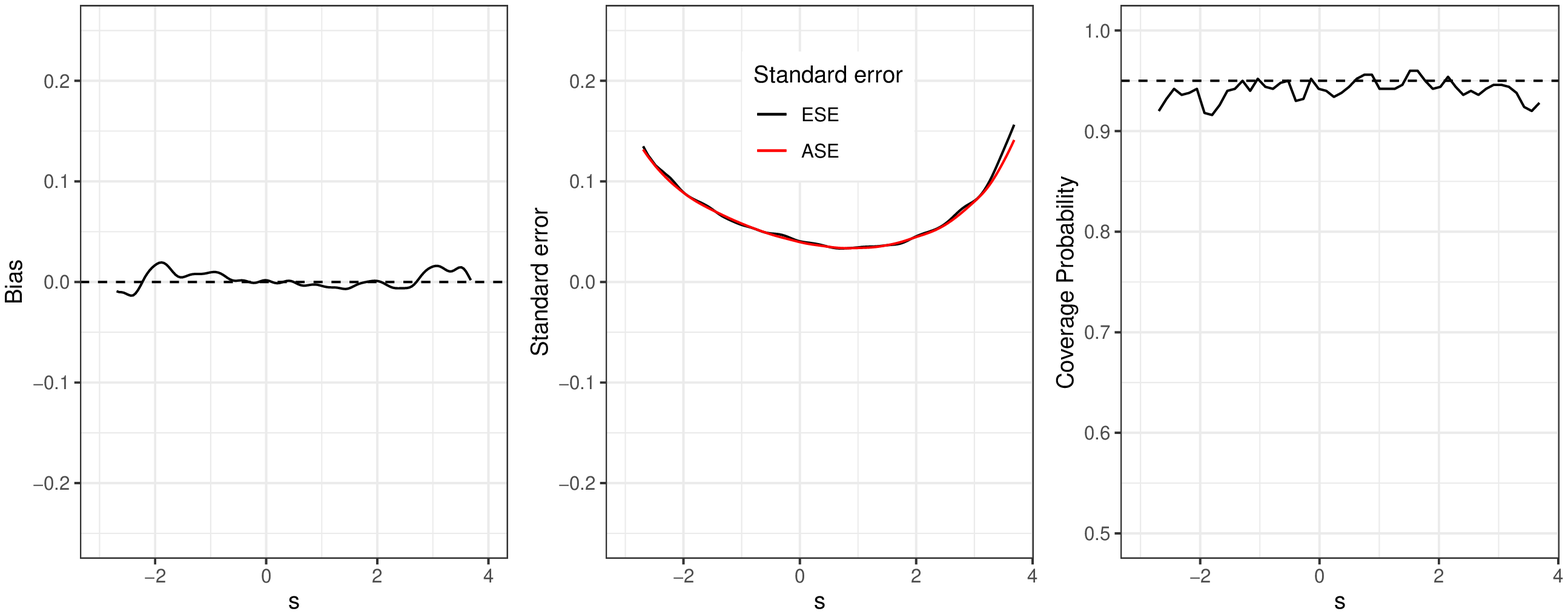}
    
  \includegraphics[scale=0.5]{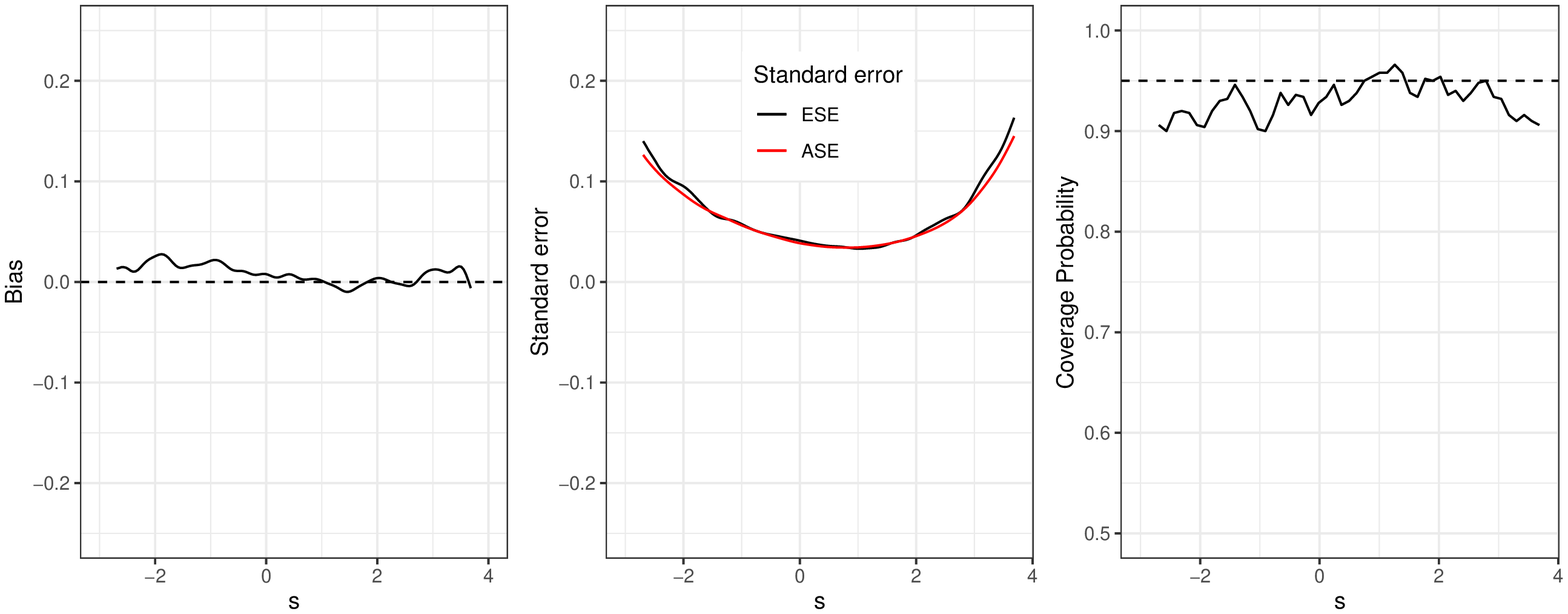}
    
  \includegraphics[scale=0.5]{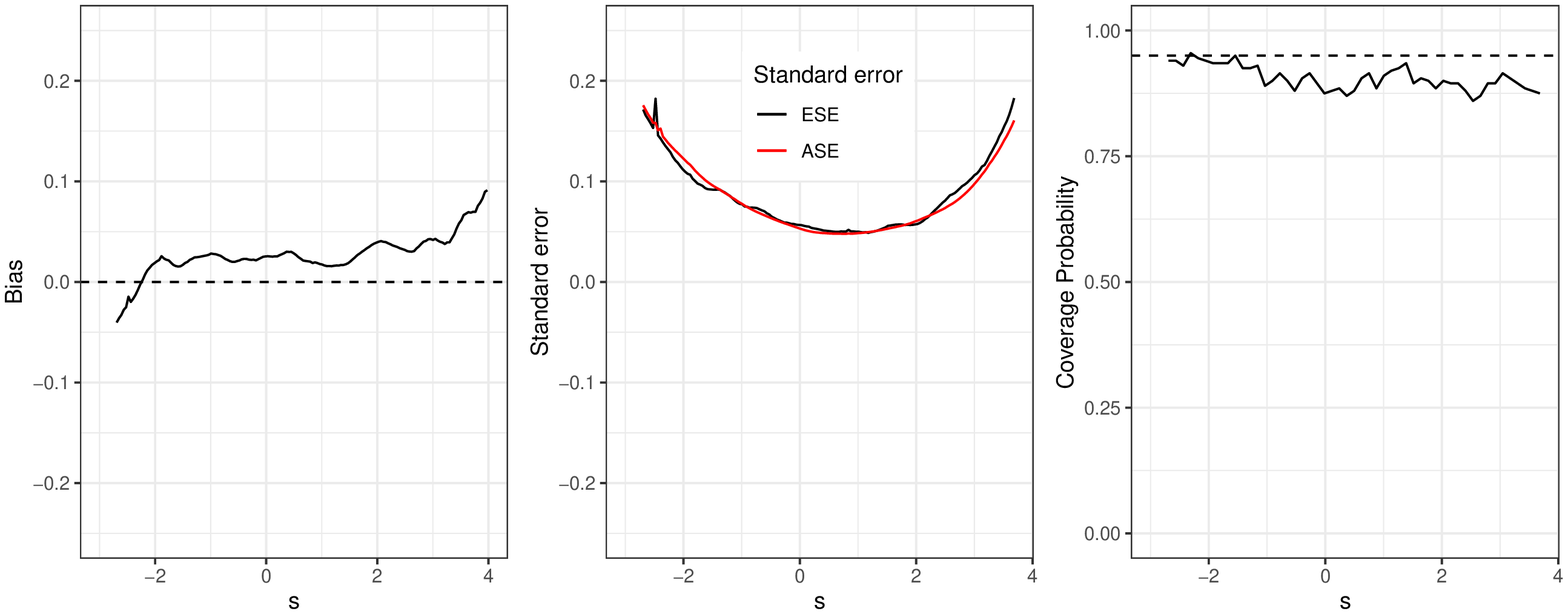}
    \caption{Empirical bias, empirical standard error (ESE) versus average of the estimated standard error (ASE), and coverage probabilities of the 95\% confidence intervals for $\ghat(s)$ when $n = 1000$ and (Row 1) both models are correctly specified, (Row 2) PS model is misspecified, (Row 3) OR model is misspecified, (Row 4) both models are misspecified.} 
    \label{fig1-4cases}
    \end{center}
\end{figure}

In Table 1, we summarize results for PTE estimation obtained via the proposed method and other existing methods. When at least one of the PS and OR models are correctly specified, the DR estimator displays negligible bias and nominal coverage. The IPW estimator $\PTEhat\subghat$ has substantial bias when the PS model is misspecified and the DR estimator also presents bias when both models are incorrect, as expected. In addition, the IPW estimator is less efficient compared to the DR estimator when the PS model is correctly specified. The estimate from $\PTEhat\subW$, which assumes that treatment is randomly assigned, shows considerable bias and below nominal coverage. It is difficult to make a direct comparison of our estimators with other literature estimators, as they estimate slightly different $PTE$ quantities, but it is useful to note that $\PTEhat\subFbX$ increases slightly from $\PTEhat\subFnaive$ toward the true $PTE$ when baseline covariates are included correctly in the OR models $Y \mid A, S, \bX$ and $Y \mid A, \bX$. However, when the OR models are misspecified, $\PTEhat\subFbX$ shows substantial bias. 

\begin{table}
	\centering
	\caption{Average of the estimated PTE, Bias for PTE estimators whose target parameter is $\PTE\subgopt=0.539$, Empirical Standard Error (ESE), Average of the Estimated Standard Errors (ASE), Root Mean Squared Error (RMSE),  and Empirical Coverage Probabilities of the $95\%$ CIs of Estimators under Different Model Scenarios for setting 1.} 
		\begin{small}
	\begin{tabular}{rlrrrrrrr}
		\hline
		Size & Estimator & Scenario & Est PTE & Bias & ESE & ASE & RMSE & Coverage \\
		\hline
		n = 400 &  $\PTEhat\subFnaive$ & No $\bX$ & 0.380 & - & 0.114 & - &  - & - \\
		&  $\PTEhat\subFbX$ & OR Correct & 0.436 & - & 0.072 & - & - & - \\
		&   & OR Misspecified & 0.611 & - & 0.122 & - & - & -  \\
		&  $\PTEhat\subFIPW$ & PS Correct & 0.460 & - & 0.203  & - & - & - \\
		&   & PS Misspecified & 0.434 & - & 0.201 & - & -  & - \\
		&  $\PTEhat\subP$ & No $\bX$  & 0.316 & - & 0.101 &  - & - & - \\ \cline{2-9}
		& $\PTEhat\subW$ & No $\bX$ & 0.435 & -0.105 & 0.110 & 0.110 & 0.191  & 0.729 \\ \cline{2-9}
		 & $\PTEhat\subghat$ & PS Correct & 0.545 & 0.006 & 0.087 & 0.088 & 0.088  & 0.930 \\
		  &  & PS Misspecified & 0.480 & -0.059 & 0.107 & 0.109 & 0.122  & 0.916 \\ \cline{2-9}
		& $\PTEhat\subghatDR$ & Both Correct & 0.542 & 0.003 & 0.079 & 0.079 & 0.080 &  0.940 \\
		 &  & PS Misspecified & 0.534 & -0.005 & 0.074 & 0.079 & 0.075  & 0.954 \\
		&  &  OR Misspecified  & 0.532 & -0.007 & 0.084 & 0.082 & 0.085  & 0.940  \\
		 &  & Both Misspecified &  0.448 & -0.091 & 0.119 & 0.115 & 0.149  & 0.779 \\

		\hline
		n = 1000 & $\PTEhat\subFnaive$ & No $\bX$ & 0.388 & - & 0.064 & - & - & -  \\
		&  $\PTEhat\subFbX$ & OR Correct & 0.445 & - & 0.040 & - & - & - \\
		&  & OR Misspecified & 0.601 & - & 0.065 & - &  - & - \\
		&  $\PTEhat\subFIPW$ & PS Correct & 0.384 & - & 0.164 & - & - & -  \\
		&   & PS Misspecified & 0.372 & - & 0.159 & - & - & - \\
		&   $\PTEhat\subP$ & No $\bX$  & 0.335 & - & 0.061  & - & - & - \\ \cline{2-9}
		 & $\PTEhat\subW$ & No $\bX$ & 0.437 & -0.103 & 0.181 & 0.174 & 0.240  & 0.803 \\ \cline{2-9}
		 	 & $\PTEhat\subghat$ & PS Correct & 0.534 & -0.006 & 0.052 & 0.053 & 0.051  & 0.950 \\
		 & & PS Misspecified & 0.453 & -0.087 & 0.069 & 0.072 & 0.113  & 0.768 \\ \cline{2-9}
		& $\PTEhat\subghatDR$ & Both Correct & 0.533 & -0.006 & 0.051  & 0.050 & 0.052  & 0.944 \\
		 &  & PS Misspecified  & 0.536 & -0.003 & 0.050 & 0.050 & 0.050 &  0.956 \\
		 &  &  OR Misspecified & 0.540 & 0.001 & 0.050 & 0.052 & 0.051  & 0.948 \\
		 &  & Both Misspecified &  0.432 & -0.107 & 0.074 & 0.072 & 0.114 & 0.635 \\
	
		\hline
	
	\end{tabular}
	\end{small}
\end{table}

\pagebreak

Table 2 shows that under setting 2, $\PTEhat\subghat$ is consistent when the PS model is correctly specified and $\PTEhat\subghatDR$ is consistent when either the PS model or OR model is correctly specified. However, $\PTEhat\subFnaive$, $\PTEhat\subFbX$, $\PTEhat\subFIPW$, and $\PTEhat\subP$ all estimate the true PTE as being close to $0$. This is in part due to the nonmonotone relationship between $Y$ and $S$ and the fact that these estimators use $S$ directly rather than $g(S)$ in estimating the treatment effect.

\begin{table}
	\centering
	\caption{Estimated PTE, Bias for PTE estimators whose target parameter is $\PTE\subgopt=0.214$, Empirical Standard Error (ESE), Average of the Estimated Standard Errors (ASE), Root Mean Squared Error (RMSE),and and Empirical Coverage Probabilities of the $95\%$ CIs of Estimators under Different Model Scenarios for setting 2.}  
	\begin{small}
	\begin{tabular}{rcrrrrrrr}
		\hline
		Size & Estimator & Scenario & Est PTE & Bias & ESE & ASE & RMSE & Coverage \\
		\hline
		
		n = 400 &  $\PTEhat\subFnaive$ & No $\bX$ & 0.048 & - & 0.031 & - &  - & - \\
		&  $\PTEhat\subFbX$ & OR Correct & 0.063 & - & 0.051 & - & - & - \\
		&   & OR Misspecified & 0.062 & - & 0.052 & - & - & -  \\
		&  $\PTEhat\subFIPW$ & PS Correct & 0.142 & - & 0.188  & - & - & - \\
		&   & PS Misspecified & 0.155 & - & 0.220 & - & -  & - \\
		&  $\PTEhat\subP$ & No $\bX$  & 0.035 & - & 0.044 &  - & - & - \\ \cline{2-9}
		& $\PTEhat\subW$ & No $\bX$ & 0.199 & -0.015 & 0.042 & 0.041 & 0.042  & 0.980 \\ \cline{2-9}
		 & $\PTEhat\subghat$ & PS Correct & 0.220 & 0.006 & 0.050 & 0.048 & 0.050  & 0.936 \\
		  &  & PS Misspecified & 0.237 & 0.024 & 0.053 & 0.052 & 0.056  & 0.894 \\ \cline{2-9}
		& $\PTEhat\subghatDR$ & Both Correct & 0.216 & 0.002 & 0.048 & 0.049 & 0.049 &  0.946 \\
		 &  & PS Misspecified & 0.214 & 0.000 & 0.047 & 0.046 & 0.048  & 0.952 \\
		&  &  OR Misspecified  & 0.219 & 0.005 & 0.048 & 0.046 & 0.048  & 0.940  \\
		 &  & Both Misspecified &  0.197 & -0.017 & 0.058 & 0.060 & 0.061  & 0.845 \\
		 
		 \hline
		n = 1000 & $\PTEhat\subFnaive$ & No $\bX$ & 0.004 & - & 0.005 & - & - & -  \\
		&  $\PTEhat\subFbX$ & OR Correct & 0.029 & - & 0.009 & - & - & - \\
		&  & OR Misspecified & 0.028 & - & 0.009 & - & - & - \\
		&  $\PTEhat\subFIPW$ & PS Correct & 0.153 & - & 0.182 & - & - & - \\
		&   & PS Misspecified & 0.141 & - & 0.157 & - & - & - \\
		&   $\PTEhat\subP$ & No $\bX$  & -0.016 & - & 0.104 & - & - & -  \\ \cline{2-9}
		 & $\PTEhat\subW$ & No $\bX$ & 0.202 & -0.011 & 0.032 & 0.030 & 0.031  & 0.970 \\ \cline{2-9}
		 	 & $\PTEhat\subghat$ & PS Correct & 0.210 & -0.004 & 0.030 & 0.030 & 0.031  & 0.948 \\
		 & & PS Misspecified & 0.232 & 0.020 & 0.030 & 0.031 & 0.031  & 0.914 \\ \cline{2-9}
		& $\PTEhat\subghatDR$ &  Both Correct & 0.218  & 0.005 & 0.028 & 0.029  & 0.029   & 0.942 \\
		 &  & PS Misspecified & 0.217 & 0.004 & 0.029 & 0.029 & 0.030   & 0.946  \\
		 &  & OR Misspecified  & 0.220 & 0.007 & 0.029 & 0.028 & 0.028 &  0.940  \\
		 &  & Both Misspecified & 0.189  & -0.024 & 0.051 & 0.045 & 0.057 & 0.853 \\
	
		\hline
	
	\end{tabular}
	\end{small}
\end{table}

\pagebreak

\section{Data Application}

Published randomized trials have shown that the partial Mayo score may be a good surrogate for the full Mayo score in assessing biologic therapies for ulcerative colitis (UC) (Lewis et al.,2008; Colombel et al., 2011; Ananthakrishnan et al., 2016). The partial Mayo score is an inexpensive, non-invasive composite score that can be measured early and ranges from $0$ to $9$. It is based on a patient's self-assessed stool frequency (0-3), rectal bleeding (0-3), and a physician's global assessment (0-3). The full Mayo score ranges from $0$ to $12$ and consists of the partial Mayo score, in addition to an invasive endoscopoy score evaluating mucosal appearance (0-3), and is typically collected later in the trial. When conventional treatments fail, biologic therapies such as infliximab, adalimumab, or golimumab may be used (Rubin et al., 2019). While these medications have been shown to be effective in placebo-controlled trials for rheumatoid arthritis (Taylor et al., 2017), there has been a lack of trials comparing agents directly in UC. One of the first such trials, a phase 3b trial of vedolizumab vs. adalimumab for patients with moderate-to-severe UC, showed that vedolizumab was superior to adalimumab with respect to achievement of clinical remission and endoscopic improvement, but not corticosteroid-free clinical remission (Sands et al., 2019). The researchers were unable to postulate an explanation for the inconsistency of the results between the primary and secondary remission outcomes, and concluded that this question required further investigation.

\begin{figure}
    \centering
    \includegraphics[scale=0.7]{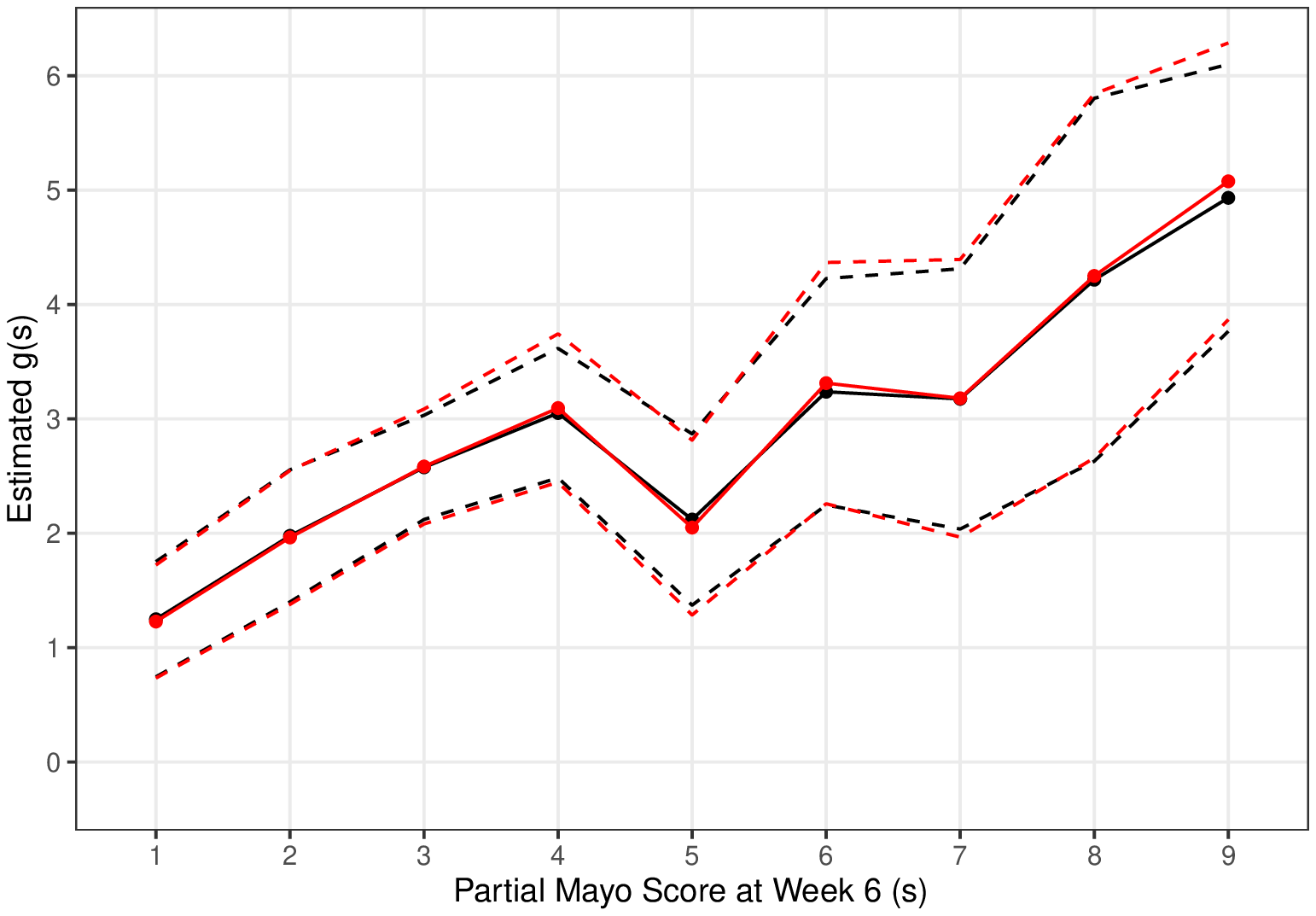}
    \caption{Estimated $g(s)$ based on IPW (red) and DR (black) estimators and pointwise 95\% confidence intervals for the partial Mayo score at week 6 (surrogate) in a cross-trial comparison of infliximab and golimumab for $361$ UC patients}
    \label{fig-example-g}
\end{figure}

To illustrate the utility of our proposed methods, we apply our procedure to examine the surrogacy of the partial Mayo score at week 6 on the primary outcome of the full Mayo score at week 54 among patients with moderate-to-severe UC. We examine an application of real-world interest in comparing head-to-head trials of two biologic therapies for patients with UC  (Ungaro \& Colombel, 2017). Treatment randomization is broken by combining data from two separate trials on patients with active UC, one comparing infliximab against a placebo (NCT00036439) and another comparing golimumab against a placebo (NCT00488631). To adjust for confounding bias, we consider baseline covariates $\bX$ including patient age, sex, race, and a health status score ranging from 0 to 100. Data was obtained from the Yale University Open Data Access (YODA) database (Ross et al., 2018). 
The ranges of $S$ in the two treatment groups are $\{1,2,...,9\}$, although the distributions are somewhat different, as evidenced in Figure 3 (see Supplementary Materials). The distributions of the primary outcome in the two treatment groups is provided in Figure 4 (Supplementary Materials). The analysis focused on the $381$ patients who had complete information on the partial Mayo score at week 6, the full Mayo score at week 54, and baseline covariates, with $216$ patients in the golimumab group and $165$ in the infliximab group. We applied the proposed methods to examine $g_{opt}(\cdot)$ of the surrogate for predicting the treatment response as quantified by the full Mayo score. The estimated $g_{opt}(\cdot)$ along with point-wise CIs based on the IPW (red) and DR (black) estimators are very similar. The estimated transformation function appears to be slightly non-linear, although there is clearly a positive trend between $s$ and $\ghat_{opt}(s)$, as shown in Figure \ref{fig-example-g}. 

The DR estimator for the treatment effect is estimated as $\widehat{\Delta} = 2.326$ in favor of golimumab and the corresponding treatment effect on the predicted outcome $\Delta\subgopt$ is estimated to be $\widehat{\Delta}\subghatDR = 2.031$. This results in a DR PTE estimate of $0.872$ with a $95\%$ CI of  $(0.741,1.003)$, suggesting that the partial Mayo score at week 6 is a strong surrogate for the full Mayo score at week 54. The results for the IPW PTE estimate is very similar at $0.878$ with a $95\%$ CI of $(0.728,1.029)$.

\section{Discussion}
There is great interest in leveraging RWD, including EHRs, registry data, and cross-trial data, to inform the design of shorter and cheaper clinical trials through surrogate marker validation. We propose the first IPW and DR estimators for the PTE explained by a surrogate marker when treatment is not randomly assigned. We generalize the approach detailed in Wang et al. (2020) for RCT data to RWD in the presence of treatment by indication bias. Our proposed doubly robust estimator is efficient and consistent when at least one of the PS and OR models is correctly specified.  In the case of UC, we have validated a partial Mayo score at week 6, which does not require an invasive endoscopy procedure, as a strong surrogate for the full Mayo score at week 54 in a cross-trial study, supplementing evidence from previous placebo-controlled trials (Lewis et al., 2008; Colombel et al., 2011; Ananthakrishnan et al., 2016). This finding may be particularly useful in informing future cross-trial designs for biologic therapies.

To provide flexibility in the estimation of $g_{opt}(\cdot)$ and $PTE$, we use a varying-coefficient model to estimate the conditional mean of $Y_i^{(a)} \mid S_i^{(a)}, \bX_i$. We are able to handle multiple confounders by implementing a two-step estimator that first reduces potentially high-dimensional $\bX$ into $\bX \trans \boldsymbol{\hat{\gamma}_a}$ through the generalized regression model and then estimates the conditional density of $S \mid \bX \trans \boldsymbol{\hat{\gamma}_a}$ using the method of Hall et al. (2004). This procedure may be computationally intensive, and in the case when the researcher is confident in the specification of the PS model, it may be advisable to consider the IPW estimation procedure. 

Our approach has some limitations. First, our proposed plug-in estimators for PTE use the same data to estimate both $g_{opt}$ and PTE given $g$, which may result in overfitting bias. However, in simulation studies, the bias appears small compared to the standard error, even with modest sample sizes. For small sample sizes, cross-validation may be needed, in which separate data is used to estimate $g_{opt}$ and PTE given $g$. Second, our approach relies on a few assumptions, the strongest of which is the working independence assumption $(Y^{(1)}, S^{(1)}) \perp (Y^{(0)},S^{(0)})$ needed for deriving the form of $\gopt$. This assumption has been discussed extensively in Wang et al. (2020). Here, we reiterate that the assumption is only a working assumption that allows for derivation of the specific form of $g_{opt}(\cdot)$ and is not required for valid inference. When the working independence assumption is severely violated, our proposed $g_{opt}$ can still be considered an optimal transformation of the surrogate marker for the difference in the primary outcome for two independent patients, one in the treatment group and the other in the control group. Third, we fit our PS models using logistic regression with specified basis functions, but alternative approaches like gradient boosting, super learner, and other machine learning classifiers may be considered in future research (Parast \& Griffin, 2017). 

\backmatter


\section*{Acknowledgements}
This study, carried out under YODA Project $\#$ 2019-4092, used data obtained from the Yale University Open Data Access Project, which has an agreement with JANSSEN RESEARCH $\&$ DEVELOPMENT, L.L.C.. The
interpretation and reporting of research using this data are solely the responsibility of the authors
and does not necessarily represent the official views of the Yale University Open Data Access Project or JANSSEN RESEARCH $\&$ DEVELOPMENT, L.L.C.. Larry Han was supported by the Harvard Big Data Training Grant (NIH-funded T32) and the Clinical Orthopedic and Musculoskeletal Education and Training (COMET) Program (NIH-funded T32) housed at Brigham and Women's Hospital, Harvard Medical School, and Harvard T.H. Chan School of Public Health.

\section*{Supplementary Materials}
\label{SM}
Supplementary material available online includes supplementary figures, additional simulation results, and R code. \vspace*{-8pt}

\begin{figure}
    \centering
    \includegraphics[scale=0.6]{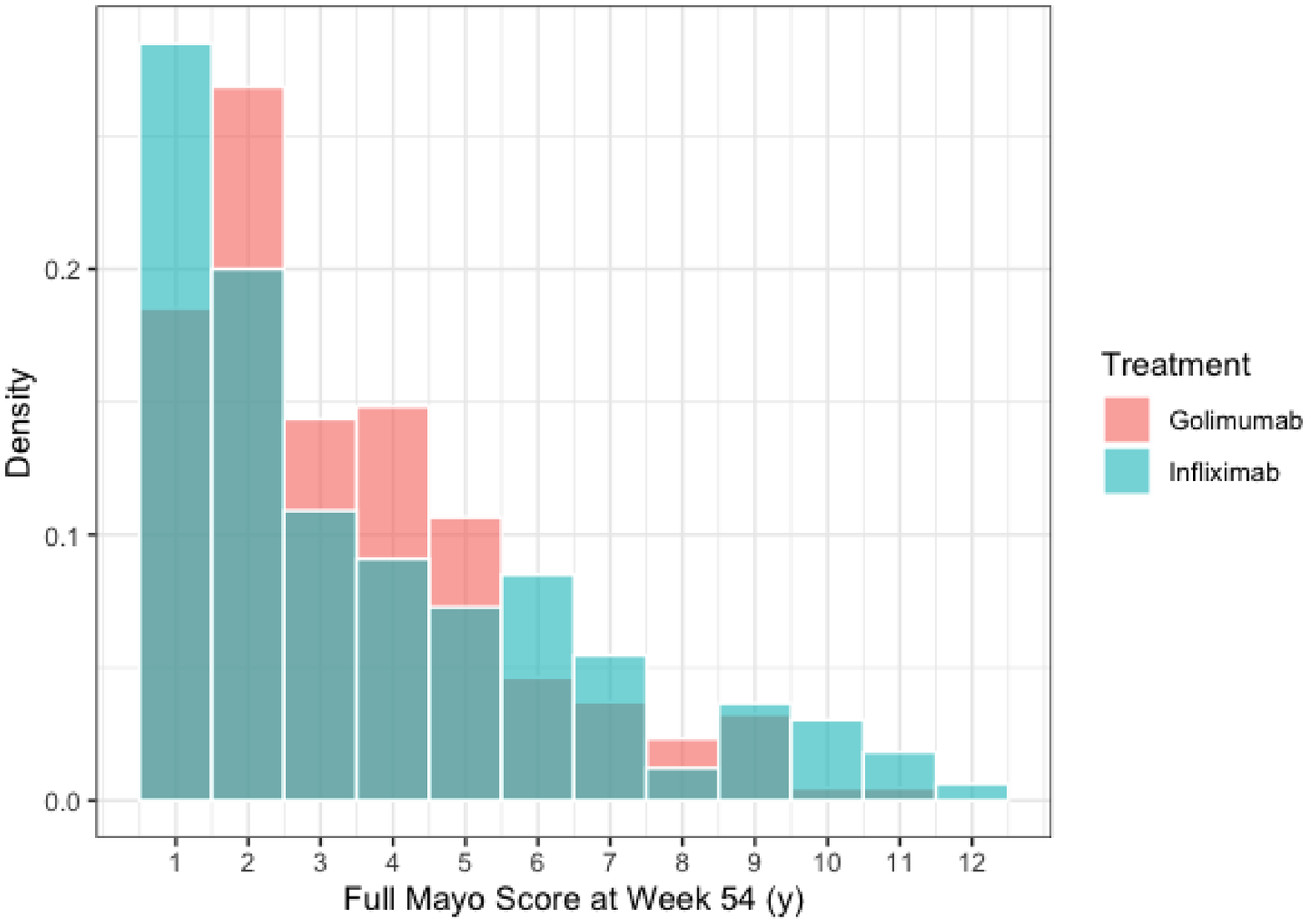}
    \caption{Histogram of the full Mayo score at week 54 (primary outcome) in the two treatment groups}
    \label{supp-fig1}
\end{figure}

\begin{figure}
    \centering
    \includegraphics[scale=0.6]{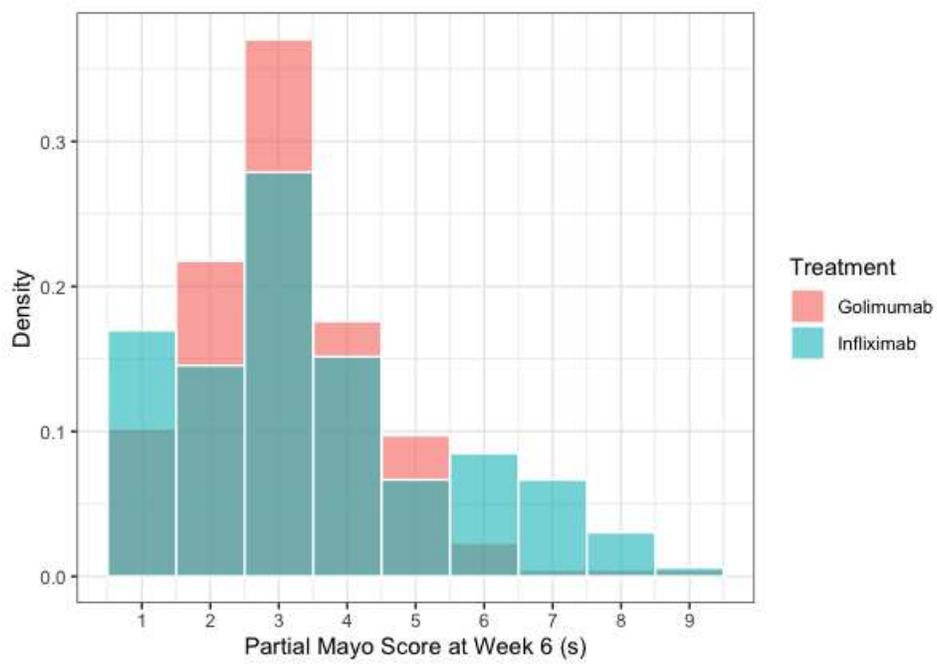}
    \caption{Histogram of the partial Mayo score at week 6 (surrogate) in the two treatment groups}
    \label{supp-fig2}
\end{figure}



\appendix
\section*{Appendix 1}
\subsection*{Consistency and asymptotic normality of $\PTEhat\subghat$}

Throughout, we assume that all components of $(Y, S, A, \bX)$ are sub-gaussian, the true conditional mean function $\psi\subam \supdag(s;\bx) = E(Y\supa \mid S\supa=s, \bX=\bx)$ and the true conditional density of $S\supa \mid \bX = \bx$, $\psi\subaf\supdag(s; \bx)$, are continuously differentiable. We also assume that $S\supa$ has a finite support and that $h=O(n^{-\nu})$ with $\nu \in (1/4,1/2)$.
In this section, we show that when the propensity score model is correctly specified, the proposed IPW kernel smoothed estimators $\ghat(s)$ and $\PTEhat\subghat$ are consistent for $\gopt(s)$ and $\PTE\subgopt$, respectively. We will also show that $\sqrt{n}(\PTEhat\subghat -\PTE\subgopt)$ converges in distribution to a normal distribution with mean zero and variance $\sigma^2$, which we will derive. To this end, we first show that $\mhat_a(s)$ and $\fhat_a(s)$ are consistent for $m_a(s)$ and $f_a(s)$, respectively. Without loss of generality, we prove the consistency of $\mhat_a(s) \equiv \mhat_a(s; \balphahat)$ for $m_a(s) = E(Y^{(a)} \mid S^{(a)}=s) = E\{\psi\subam\supdag(s; \bX)\}$, where
$$\mhat_a(s; \balpha) = \frac{n^{-1}\sum_{i=1}^n K_h(S_i-s)Y_iI(A_i=a)/\pi_a(\bX_i; \balpha)}{n^{-1}\sum_{i=1}^n K_h(S_i-s)I(A_i=a)/\pi_a(\bX_i; \balpha)} .$$
First, under the correct specification of the PS model, $\balphahat \to \balpha_0$ in probability, where $\balpha_0$ is the true parameter value. Hence $\max_i | \omegahat_{ai} - \omega_{ai}| \le \sup_{\bx} |\pi_a(\bx; \balphahat)^{-1}-\pi_a(\bx; \balpha_0)^{-1}| \to 0$ in probability, where $\omega_{ai}=I(A_i=a)/\pi_a(\bX_i; \balpha_0)$. It then follows from standard theory for non-parametric kernel estimators (Masry, 1996; Pagan \& Ullah, 1999)  and Taylor series expansions that 
\begin{align*}
 \sup_s|\mhat_a(s; \balphahat)-m_a(s)| \le & \sup_s |\mhat_a(s; \balpha_0)-m_a(s)| + \sup_{s, \balpha: \|\balpha-\balpha_0\|\le c}\|\bmhat'_a(s;\balpha)\|_2 \|\balphahat - \balpha_0\|_2 \\
= & O_p\{(nh)^{-\half}\sqrt{\log n} + h^2 +\nnhalf\}=o_p(1)    
\end{align*} where $\bmhat_a'(s;\balpha)=\partial \mhat_a(s; \balpha)/\partial \balpha$ and $c$ is any small constant. Similarly, we have $\sup_s|\fhat_a(s)-f_a(s)|= O_p\{(nh)^{-\half}\sqrt{\log n} + h^2 +\nnhalf\}=o_p(1)$. When $h=O(n^{-\nu})$ with $\nu \in (1/4,1/2)$, it is not difficult to show that $\lambdahat - \lambda = O_p(\nnhalf + h^2) = O_p(\nnhalf)$. It follows that $\sup_s|\ghat(s)-g(s)|= O_p\{(nh)^{-\half}\sqrt{\log n} + h^2 +\nnhalf\}=o_p(1)$. Similarly, we may show that 
\begin{align*}
    & |\Deltahat\subghat - \Delta\subgopt| = \Deltahat\subghat - \Deltahat\subgopt + \Deltahat\subgopt - \Delta\subgopt 
    = \int \{\ghat(s)-\gopt(s)\} d \Dhat(s)  + O_p(\nnhalf),
\end{align*} 
where $\Dhat(s) = \ninv\sumin  (\omegahat_{1i} - \omegahat_{0i}) I(S_i\le s)$. It follows from the uniform convergence of $\ghat(s) \to \gopt(s)$ and $\Dhat(s)\to D(s) = P(S\supone \le s)-P(S\supzero \le s)$ that $\Deltahat\subghat - \Delta\subgopt \to 0$ in probability. This, together with the consistency of $\Deltahat$ for $\Delta$, implies the consistency of $\PTEhat\subghat$ for $\PTE\subgopt$.

We next establish the asymptotic normality of $\sqrt{n}(\PTEhat\subghat-\PTE\subgopt)$. 
First, note that
\begin{align*}
\fhat_{a}(s)-f_{a}(s) &= n^{-1} \sum_{i=1}^{n} \left[\omega_{ai} \left\{K_{h}\left(S_{i}-s\right)-f_{a}(s)\right\}+ \bbf'_a(s)\bUsc_{\alpha,i}\right] + o_p((nh)^{-1/2}), \\
\widehat{m}_a(s)-m_a(s) &= n^{-1} \sum_{i=1}^{n}\left[ \omega_{ai} K_{h}\left(S_{i}-s\right) \mathcal{U}_{m_a,i}(s)+\bm'_a(s)\bUsc_{\alpha,i}\right] + o_p((nh)^{-1/2}),
\end{align*}
where $\bm_a'(s; \balpha) = \partial m_a(s; \balpha)/\partial \balpha$, $\bbf'_a(s; \balpha) = \partial f_a(s; \balpha)/\partial \balpha$, 
$\mathcal{U}_{m_a,i}(s) = f_a(s)^{-1}\{Y_i^{(a)}-m_a(s)\}$,
\begin{align*}
    m_a(s; \balpha) & = \frac{E\left\{\psi\subam\supdag(s;\bX_i)\psi\subaf\supdag(s;\bX_i)\frac{\pi(\bX_i; \balpha_0)}{\pi(\bX_i; \balpha)} \right\}}{f_a(s; \balpha)}, \quad 
    f_a(s; \balpha)  = E\left\{\psi\subaf\supdag(s;\bX_i)\frac{\pi(\bX_i; \balpha_0)}{\pi(\bX_i; \balpha)}\right\} ,
\end{align*} 
and $\balphahat - \balpha_0 = \ninv\sumin \bUsc_{\alpha,i}+o_p(\nnhalf)$ following standard likelihood theory.
It follows that
 
\begin{equation*}
\begin{aligned}
&\widehat{\mathcal{P}}_{0}(s)-\mathcal{P}_{0}(s) =\frac{\hat{f}_{0}(s) f_{1}(s)-\hat{f}_{1}(s) f_{0}(s)}{\left\{f_{1}(s)+f_{0}(s)\right\}^{2}}+o_p((nh)^{-1/2}) \\
&= \frac{\Psc_1(s)\Psc_0(s)(\hat{f}_0(s)-f_0(s))}{f_0(s)} - \frac{\Psc_1(s)\Psc_0(s)(\hat{f}_1(s)-f_1(s))}{f_1(s)} + o_p((nh)^{-1/2})\\
&= \Psc_1(s)\Psc_0(s) \ninv \sumin \left[ \omega_{0i}\bigg\{K_h(S_i-s)f_0(s)^{-1}-1\right\} - \omega_{1i}\left\{K_h(S_i-s) f_1(s)^{-1}-1\right\} \\ &+ \bUsc_{\alpha,i}\trans \left\{ \frac{\bbf'_0(s)}{f_0(s)}-\frac{\bbf'_1(s)}{f_1(s)} \right\} \bigg] + o_p((nh)^{-1/2}) \\
&= \ninv \sumin \left[K_h(S_i-s) \Gsc_{\Psc_0,i}(s) + (\omega_{1i}-\omega_{0i})\Psc_{1}(s)\Psc_0(s)  + \bUsc_{\alpha,i}\trans\bBsc_{\Psc_0}(s)\right] + o_p((nh)^{-1/2}) \\
&= \ninv\sumin \Usc_{\Psc_0,i}+o_p((nh)^{-1/2}),
\end{aligned}
\end{equation*}
where
\begin{align*}
\Usc_{\Psc_0,i}(s)
&=  K_h(S_i-s) \Gsc_{\Psc_0,i}(s) + (\omega_{1i}-\omega_{0i})\Psc_{1}(s)\Psc_0(s)  + \bUsc_{\alpha,i}\trans\bBsc_{\Psc_0}(s), \\
 \Gsc_{\Psc_0,i}(s) &= \{\omega_{0i}/f_0(s) - \omega_{1i}/f_1(s)\}\Psc_1(s)\Psc_0(s), \\
\bBsc_{\Psc_0}(s) &= \{\bbf'_0(s)/f_0(s)-\bbf'_1(s)/f_1(s)\}\Psc_1(s)\Psc_0(s) 
\end{align*}

Similarly, we have $\Pschat_1(s)-\Psc_1(s) = -(\Pschat_0(s)-\Psc_0(s)) = -\ninv\sumin \Usc_{\Psc_0,i}(s)+o_p(\nnhalf)$.

Now 
\begin{align*}
    \mhat(s)-m(s) &= \{\mhat_1(s)-m_1(s)\}\Psc_1(s) + \mhat_1(s)\{\Pschat_1(s)-\Psc_1(s)\} \\
    &\quad+ \{\mhat_0(s)-m_0(s)\}\Psc_0(s) + \mhat_0(s)\{\Pschat_0(s)-\Psc_0(s)\} \\
    &= \ninv\sumin \Usc_{m,i}(s) + o_p((nh)^{-1/2}),
\end{align*}
where
\begin{equation*}
\begin{aligned}
\Usc_{m,i}(s) &= \sum_{a=0}^1 \left\{\omega_{ai}K_h(S_i-s)\Usc_{m_a,i}(s) + \bm_a'(s)\bUsc_{\alpha,i} \right\}\Psc_a(s) + \{m_0(s)-m_1(s)\}\Usc_{\Psc_0,i}(s) \\
&=K_h(S_i-s) \Gsc_{m,i}(s) + \bUsc_{\alpha,i}\trans\bBsc_{m}(s) + (\omega_{1i}-\omega_{0i}) \Asc_{m}(s),
\end{aligned}
\end{equation*}
where
\begin{align*}
    \Gsc_{m,i}(s) &= \sum_{a=0}^1 \omega_{ai} \Usc_{ma,i}(s) \Psc_a(s) +  \Gsc_{\Psc_0,i}(s)\{m_0(s)-m_1(s)\},\\
    \bBsc_{m}(s) &= \sum_{a=0}^1 \bm_a'(s)\Psc_a(s) + \bBsc_{\Psc_0}(s)\{m_0(s)-m_1(s)\}, \\
    \Asc_m(s) &=  \Psc_1(s)\Psc_0(s)\{m_0(s)-m_1(s)\}
\end{align*}

Together with arguments given in Appendix D of Wang et al. (2020), the fact that $h = o_p(n^{-1/4})$, and a Taylor series expansion for approximating $\int K_h(S_i-s)H(s)ds$ for any given smooth function $H$, where we denote $\mu_{H_0} := \int H(s) dF_0(s)$, we have the following expansion for $\int  \widehat{m}(s) d \widehat{F}_{0}(s)-\mu_{m 0}$:
\begin{equation*}
\begin{aligned}
    \int  \widehat{m}(s) \fhat_{0}(s)ds -\mu_{m 0}&=\int\{\widehat{m}(s)-m(s)\} f_0(s) ds \\ &\quad+\int m(s) \left\{\fhat_{0}(s)-f_{0}(s)\right\}ds + o_p(\nnhalf) \\
    &= \ninv\sumin \bigg[\int K_h(S_i-s)\left\{\Gsc_{m,i}(s) f_0(s) + \omega_{0i}m(s)\right\} ds \\
    &\quad+ (\omega_{1i}-\omega_{0i})\int \Asc_{m}(s) f_0(s)ds -\omega_{0i}\int m(s)f_0(s)ds \\
    &\quad+ \bUsc_{\alpha,i}\trans\int \left\{m(s)\bbf'_0(s) +\bBsc_m(s) f_0(s)\right\}ds\bigg] + o_p(n^{-1/2}) \\
    &=\ninv\sumin \bigg[\Gsc_{m,i}(S_i) f_0(S_i) + \omega_{0i}m(S_i) + (\omega_{1i}-\omega_{0i})\int \Asc_{m}(s) f_0(s)ds\\&\quad -\omega_{0i}\int m(s)f_0(s)ds + \bUsc_{\alpha,i}\trans\int \left\{m(s)\bbf'_0(s) +\bBsc_m(s) f_0(s)\right\}ds\bigg] \\
    &\quad+ o_p(n^{-1/2}). 
\end{aligned}
\end{equation*}
Similarly, we have
\begin{equation*}
    \begin{aligned}
    \int \widehat{P}_{0}(s) &\fhat_{0}(s) ds-\mu_{ \mathcal{P}_00}\\
    &=\int\left\{\widehat{P}_{0}(s)-\mathcal{P}_{0}(s)\right\} f_{0}(s)ds +\int \mathcal{P}_{0}(s) \left\{\fhat_{0}(s)-f_{0}(s)\right\} ds+ o_p(n^{-1/2}) \\
    &=\ninv\sumin \bigg[ \int K_h(S_i-s)\{\Gsc_{\Psc_0,i}(s) f_0(s) + \omega_{0i} \Psc_0(s)\}ds \\
    &\quad+ (\omega_{1i}-\omega_{0i})\int \Psc_1(s)\Psc_0(s) f_0(s) ds - \omega_{0i} \int \Psc_0(s) f_0(s) ds \\
    &\quad+ \bUsc_{\alpha,i}\trans\int \left\{\Psc_0(s)\bbf'_0(s) +\bBsc_{\Psc_0}(s) f_0(s)\right\}ds\bigg] + o_p(n^{-1/2}) \\
    &= \ninv\sumin \bigg[\Gsc_{\Psc_0,i}(S_i) f_0(S_i) + \omega_{0i}\Psc_0(S_i) + (\omega_{1i}-\omega_{0i})\int \Psc_1(s)\Psc_0(s) f_0(s) ds \\ &\quad - \omega_{0i} \int \Psc_0(s) f_0(s) ds +
    \bUsc_{\alpha,i}\trans\int \left\{\Psc_0(s)\bbf'_0(s) +\bBsc_{\Psc_0}(s) f_0(s)\right\}ds \bigg] \\&\quad+ o_p(n^{-1/2}).
     \end{aligned}
\end{equation*}

Since $\lambda = \frac{\displaystyle{\mu_0 - \mu_{m0}}}{\displaystyle{\mu_{\mathcal{P}_00}}}$ and $\widehat{\lambda}= \frac{\displaystyle{\widehat{\mu}_{0}-\widehat{\mu}_{m 0}}}{ \displaystyle{\widehat{\mu}_{\mathcal{P}_00}}}$, it follows from above that
$$\lambdahat - \lambda = \ninv\sumin \Usc_{\lambda, i}+o_p(\nnhalf),$$ where 
\begin{equation*}
    \begin{aligned}
    \widehat{\lambda}-\lambda&= \mu_{\mathcal{P}_{0} 0}^{-1}\left(\widehat{\mu}_{0}-\mu_{0}\right)-\mu_{\mathcal{P}_{0} 0}^{-1} \lambda\left\{\int \widehat{\mathcal{P}}_{0}(s) d \widehat{F}_{0}(s)-\mu_{\mathcal{P}_{0} 0}\right\}\\ &\quad-\mu_{\mathcal{P}_{0} 0}^{-1}\left\{\int \widehat{m}(s) d \widehat{F}_{0}(s)-\mu_{m 0}\right\}+ o_p(n^{-1/2}) \\
&= \mu_{\mathcal{P}_{0} 0}^{-1} n^{-1} \sum_{i=1}^{n} w_{0 i}\left(Y_{i}-\mu_{0}\right) -\mu_{\mathcal{P}_{0} 0}^{-1} \lambda  n^{-1} \sum_{i=1}^{n} \bigg\{ \Gsc_{\Psc_0,i}(S_i) f_0(S_i) + \omega_{0i}\Psc_0(S_i) \\
&\quad+ (\omega_{1i}-\omega_{0i})\int \Psc_1(s)\Psc_0(s) f_0(s) ds - \omega_{0i} \int \Psc_0(s) f_0(s) ds \bigg\} \\ 
&\quad+\mu_{\mathcal{P}_{0} 0}^{-1} \lambda n^{-1} \sum_{i=1}^{n} \bUsc_{\alpha,i}\trans\int \left\{\Psc_0(s)\bbf'_0(s) +\bBsc_{\Psc_0}(s) f_0(s)\right\}ds \\ 
&\quad- \mu_{\mathcal{P}_{0} 0}^{-1} n^{-1} \sum_{i=1}^{n}\bigg\{\Gsc_{m,i}(S_i) f_0(S_i) + \omega_{0i}m(S_i) \\
&\quad+ (\omega_{1i}-\omega_{0i})\int \Asc_{m}(s) f_0(s)ds -\omega_{0i}\int m(s)f_0(s)ds \bigg\}  \\
    &\quad+ \mu_{\mathcal{P}_{0} 0}^{-1} n^{-1} \sum_{i=1}^{n}\bUsc_{\alpha,i}\trans\int \left\{m(s)\bbf'_0(s)ds +\bBsc_m(s) f_0(s)\right\}ds + o_p(n^{-1/2})  \\
&= n^{-1}\sumin \mathcal{U}_\lambda(\textbf{D}_i) + o_p(n^{-1/2}).
    \end{aligned}
\end{equation*}

Gathering the above expansions, we may obtain the form of $\widehat{g}(s) - \gopt(s)$ as 
\begin{equation*}
    \begin{aligned}
    \widehat{g}(s)&-g_{\mathrm{opt}}(s)\\
    &=\widehat{m}(s)-m(s)+(\widehat{\lambda}-\lambda) \mathcal{P}_{0}(s)+\lambda\left\{\widehat{\mathcal{P}}_{0}(s)-\mathcal{P}_{0}(s)\right\} +o_p((nh)^{-1/2})\\
    &= n^{-1} \sumin \bigg[K_h(S_i-s) \Gsc_{m,i}(s) + \bUsc_{\alpha,i}\trans\bBsc_{m}(s) + (\omega_{1i}-\omega_{0i}) \Asc_{m}(s) +\mathcal{P}_{0}(s) \mathcal{U}_\lambda(\textbf{D}_i) \\
    &\quad+ \lambda \{K_h(S_i-s) \Gsc_{\Psc_0,i}(s) + (\omega_{1i}-\omega_{0i})\Psc_{1}(s)\Psc_0(s)  + \bUsc_{\alpha,i}\trans\bBsc_{\Psc_0}(s)\}  \bigg] \\
    &\quad+o_p((nh)^{-1/2})\\
    &= n^{-1} \sumin \mathcal{U}_G(s;\textbf{D}_i) + o_p((nh)^{-1/2}),
    \end{aligned}
\end{equation*}
where \begin{align*}
    \mathcal{U}_G(s;\textbf{D}_i) &= K_h(S_i-s)\{ \Gsc_{m,i}(s)+\lambda \Gsc_{\Psc_0,i}(s)\} + \bUsc_{\alpha,i}\trans\{\bBsc_{m}(s)+\lambda \bBsc_{\Psc_0}(s)\} \\
    &\quad+ (\omega_{1i}-\omega_{0i}) \{\Asc_{m}(s)+\lambda \Psc_{1}(s)\Psc_0(s)\} +\mathcal{P}_{0}(s) \mathcal{U}_\lambda(\textbf{D}_i) .
\end{align*}

To derive the asymptotic distribution for $\widehat{\mathrm{PTE}},$ observe that
\[
\begin{aligned}
\widehat{\mathrm{PTE}}&-\mathrm{PTE} \\
&= \int\left\{\widehat{g}(s)-g_{\mathrm{opt}}(s)\right\} d\left\{\widehat{F}_{1}(s)-\widehat{F}_{0}(s)\right\} +\int g_{\mathrm{opt}}(s) d\left\{\widehat{F}_{1}(s)-\widehat{F}_{0}(s)\right\}-\mathrm{PTE} \\
&=\frac{1}{\Delta} n^{-1} \sumin \bigg[(\Gsc_{m,i}(S_i)+\lambda \Gsc_{\Psc_0,i}(S_i))(f_1(S_i) - f_0(S_i)) + (\omega_{1i}-\omega_{0i})g_{\mathrm{opt}}(S_i) \\
&\quad+ \int g_{\mathrm{opt}}(s) \{\bbf'_1(s)-\bbf'_0(s)\}\bUsc_{\alpha,i}ds - \mathrm{PTE} + \int \bigg\{\bUsc_{\alpha,i}\trans\{\bBsc_{m}(s)+\lambda \bBsc_{\Psc_0}(s)\} \\
&\quad+ (\omega_{1i}-\omega_{0i}) \{\Asc_{m}(s)+\lambda \Psc_{1}(s)\Psc_0(s)\} +\mathcal{P}_{0}(s) \mathcal{U}_\lambda(\textbf{D}_i)\bigg\}(f_1(s)-f_0(s))ds\bigg]  \\
&\quad-\frac{\mathrm{PTE}}{\Delta} n^{-1} \sumin [\omega_{1i}(Y_i-\mu_1)-\omega_{0i}(Y_i-\mu_0)]  + o_p(n^{-1/2})\\
&= n^{-1} \sumin \mathcal{U}_{\mathrm{PTE}}(\textbf{D}_i) + o_p(n^{-1/2}),
\end{aligned}
\]
where 
\begin{equation*}
    \begin{aligned}
    \mathcal{U}_{\mathrm{PTE}}(&\textbf{D}_i) \\
    &=\frac{1}{\Delta} \bigg[ (\Gsc_{m,i}(S_i)+\lambda \Gsc_{\Psc_0,i}(S_i))(f_1(S_i) - f_0(S_i)) + (\omega_{1i}-\omega_{0i})g_{\mathrm{opt}}(S_i) \\ 
    &\quad+\int g_{\mathrm{opt}}(s) \{\bbf'_1(s)-\bbf'_0(s)\}\bUsc_{\alpha,i}ds - \mathrm{PTE} + \int \bigg\{\bUsc_{\alpha,i}\trans\{\bBsc_{m}(s)+\lambda \bBsc_{\Psc_0}(s)\} \\
    &\quad+ (\omega_{1i}-\omega_{0i}) \{\Asc_{m}(s)+\lambda \Psc_{1}(s)\Psc_0(s)\} +\mathcal{P}_{0}(s) \mathcal{U}_\lambda(\textbf{D}_i)\bigg\} (f_1(s)-f_0(s))ds\bigg]\\
&\quad-\frac{\mathrm{PTE}}{\Delta}   [\omega_{1i}(Y_i-\mu_1)-\omega_{0i}(Y_i-\mu_0)].
    \end{aligned}
\end{equation*}

Therefore, by the central limit theorem, $\sqrt{n} \left(\widehat{\mathrm{PTE}}-\mathrm{PTE}\right)$  converges in distribution to a normal with mean zero and variance $\sigma^{2}=E\left\{\mathcal{U}_{\mathrm{PTE}}\left(\mathbf{D}_{i}\right)^{2}\right\}$. 

\pagebreak

\section*{Appendix 2}
\subsection*{Double Robustness}

In this section, we prove that our proposed DR estimators are consistent when either the PS model or the OR models are correctly specified. Recall that we proposed the augmented IPW estimators for 
$m_a(s)$ and $f_a(s)$,
\begin{subequations}
\begin{align}
    \Mschat\subaDR(s) &= n^{-1} \sum_{i=1}^n \left\{K_h(S_i-s) Y_i \omegahat_{ai}- (\omegahat_{ai}-1) \psihat\supdag\subam(s; \bX_i)\psihat\supdag\subaf(s; \bX_i) \right\} , \\
    \fhat\subaDR(s)  &= n^{-1} \sum_{i=1}^n \left\{K_h(S_i-s)\omegahat_{ai} - (\omegahat_{ai}-1) \psihat\subaf(s; \bX_i)  \right\} 
    \\
    \mhat\subaDR(s) &= \frac{\Mschat\subaDR(s)}{\fhat\subaDR(s)}
\end{align}
\end{subequations}
respectively, where $h = O(n^{-\nu})$ with $\nu \in (1/4, 1/2)$, $\psihat\subam(s; \bx)$ and $\psihat\subaf(s; \bx)$ are estimators for the conditional mean $\psi\subam\supdag(s; \bx)$ and the conditional density $\psi\subaf\supdag(s; \bx)$, respectively. 

We now show that the estimators are consistent if either $\sup_{\bx}|\pi_a(\bx; \balphahat)-\pi_a(\bx)|\to 0$ in probability or $\sup_{\bx,s}\{|\psihat\subam(s; \bx)-\psi\subam\supdag(s; \bx)|+|\psihat\subaf(s; \bx)-\psi\subaf\supdag(s; \bx)|\}\to 0$ in probability. Let $\balphabar$, $\psibar\subam(s; \bx)$, $\psibar\subaf(s;\bx)$ denote the respective limits of $\balphahat$, $\psihat\subam(s; \bx)$ and $\psihat\subaf(s; \bx)$ under possible mis-specification of their respective models,   $\pibar_a(\bx) = \pi_a(\bx; \balphabar)$, and $\omegabar_{ai} = I(A_i=a)/\pibar_a(\bX_i)$. Regardless of the adequacy of the models, by the central limit theorem and convergence of kernel smoothed estimators (Pagan \& Ullah, 1999), we have that $\balphahat - \balphabar=O_p(\nnhalf)$ and $\sup_{s,\bx}|\psihat\subaf(s; \bx)-\psibar\subaf(s;\bx)|+\sup_{s,\bx}|\psihat\subam(s;\bx)-\psibar\subam(s;\bx)|=o_p(1)$.

When the PS model is correctly specified, $\sup_s|\mhat_a(s) - m_a(s)| + \sup_s|\fhat_a(s)-f_a(s)| \to 0$ in probability as shown in Appendix 1. In addition, the augmentation terms $$n^{-1} \sum_{i=1}^n (\omegahat_{ai}-1) \psihat\subam(s; \bX_i) = n^{-1} \sum_{i=1}^n (\omega_{ai}-1) \psibar\subam(s; \bX_i) + O_p(\|\balphahat-\balpha_0\|_2) $$ and $$n^{-1} \sum_{i=1}^n (\omegahat_{ai}-1) \psihat\subaf(s; \bX_i)= n^{-1} \sum_{i=1}^n (\omega_{ai}-1) \psibar\subaf(s; \bX_i) + O_p(\|\balphahat-\balpha_0\|_2)$$
also converge to 0 in probability, regardless of the adequacy of the OR models. Therefore, under the correct specification of the PS model, $\sup_s|\mhat\subaDR(s) - m_a(s)| + \sup_s|\fhat\subaDR(s)-f_a(s)| \to 0$ in probability.

We next establish the consistency of the DR estimators when the PS model may be mis-specified but the OR models are correctly specified. First consider $\fhat\subaDR(s)$, which can be written as
\begin{align*}
\fhat\subaDR(s) &= n^{-1}  \sum_{i=1}^n \bigg[ \{K_h(S_i-s)-\psi\subaf\supdag(s;\bX_i)\}\omegahat_{ai} \\
&\quad- (\omegahat_{ai}-1) \{\psihat\subaf(s; \bX_i)  -\psi\subaf\supdag(s;\bX_i) \} + \psi\subaf\supdag(s;\bX_i) \bigg] \\
&= n^{-1}  \sum_{i=1}^n \bigg[ \{K_h(S_i-s)-\psi\subaf\supdag(s;\bX_i)\}\omegahat_{ai} \\
&\quad- (\omegahat_{ai}-1) \{\psihat\subaf(s; \bX_i)  -\psi\subaf\supdag(s;\bX_i) \}  \bigg] + f_a(s)+O_p(\nnhalf) ,  \\ 
&= n^{-1}  \sum_{i=1}^n \varepsilon\subaDR(s;\bD_i)  + f_a(s) - n^{-1}  \sum_{i=1}^n  (\omegabar_{ai}-1) \{\psihat\subaf(s; \bX_i)  -\psi\subaf\supdag(s;\bX_i) \}   \\&\quad+ O_p(\nnhalf) , 
\end{align*}
where $\varepsilon\subaf(s;\bD_i) = \{K_h(S_i-s)-\psi\subaf\supdag(s;\bX_i)\}\omegabar_{ai}$. It follows from the uniform convergence of kernel smoothed estimators (Pagan \& Ullah, 1999) that  $$\sup_{s}|n^{-1}  \sum_{i=1}^n \varepsilon\subaDR(s;\bD_i) - E\{\varepsilon\subaDR(s;\bD_i)\}| = o_p(1)$$ and $$E\{\varepsilon\subaDR(s;\bD_i)\} = E\left[\omegabar_{ai}\left\{\int K_h(S-s)\psi\subaf\supdag(S;\bX_i)dS -\psi\subaf\supdag(s;\bX_i)\right\}\right] = O(h^2).$$
This together with $\sup_{s,\bx}|\psihat\subaf(s; \bx) - \psi\supdag\subaf(s; \bx)|=o_p(1)$ implies that $\sup_{s}|\fhat\subaDR(s)-f_a(s)|=o_p(1)$. We have a similar consistency result for $\Mschat\subaDR(s)$, where 

\begin{align*}
\Mschat\subaDR(s)  &= n^{-1} \sum_{i=1}^n \left\{K_h(S_i-s) Y_i \omegabar_{ai}- (\omegabar_{ai}-1) \psi\supdag\subam(s; \bX_i)\psi\supdag\subaf(s; \bX_i) \right\} + o_p(1) \\
&= n^{-1} \sum_{i=1}^n \bigg[\eps\subam(s;\bD_i) + \left\{K_h(S_i-s)\psi\subam\supdag(s; \bX_i)  - \psi\supdag\subam(s; \bX_i)\psi\supdag\subaf(s;\bX_i) \right\} \omegabar_{ai}\bigg]\\
&\quad+ m_a(s)+ o_p(1),  
\end{align*}
and $\eps\subam(s;\bD_i) = K_h(S_i-s)\{ Y_i - \psi\subam\supdag(s; \bX_i)\}\omegabar_{ai}$.
Following the convergence of $\fhat\subaDR(s) \to f_a(s)$, $\Mschat\subaDR(s) \to m_a(s)f_a(s)$, $\psihat\subam(s;\bx)\to\psi\subam\supdag(s;\bx)$ and $\omegahat_{ai}\to \omegabar_{ai}$, we arrive at the consistency of $\mhat\subaDR(s)$ to $m_a(s)$ when the PS model may be mis-specified but the OR models are correctly specified.

Thus, we get the double robustness properties for $\fhat\subaDR(s)$ and $\mhat\subaDR(s)$.

Since all remaining estimators relevant to $\ghat\subDR(s)$ are plug-in estimators that are derived based on $\mhat\subaDR(s)$ and $\fhat\subaDR(s)$, we can conclude the double robustness of $\ghat\subDR(s)$ for $\gopt(s)$. 

Finally, the PTE will be doubly robust by standard arguments for the conditional mean estimators (Robins et al., 1994), where we construct a plug-in estimator for $\Delta_{\gopt}$ as $\widehat{\Delta}\subghatDR= \muhat\suboneghatDR - \muhat\subzeroghatDR,$
where  
$$\muhat_{a,g,DR} = n_a^{-1}\sum_{i:A_i=a} \left\{ \frac{g(S_i)}{\hat{\pi}_a(\bX_i)} - \frac{I(A_i=a) - \hat{\pi}_a(\bX_i)}{\hat{\pi}_a(\bX_i)}\hat{\zeta}_{a,g}(\bX_i) \right\} ,$$
where $\hat{\zeta}_{a,g}(\bx)$ is an estimator for $\zeta_{a,g}(\bx) = E(g(S_i^{(a)}) \mid \bX_i = \bx) = E(g(S_i) \mid A_i=a, \bX_i = \bx)$, and $n_a = \sum_{i=1}^n I(A_i=a)$, $a=0,1$. Similarly, we define $\widehat{\Delta}\subDR = \muhat\suboneDR - \muhat\subzeroDR$, where $$\muhat\subaDR = n_a^{-1} \sum_{i: A_i=a} \left\{ \frac{Y_i}{\pihat_a(\bX_i)} - \frac{I(A_i=a) - \pihat_a(\bX_i)}{\pihat_a(\bX_i)}\hat{\zeta}_a(\bX_i) \right\},$$ where $\hat{\zeta}_a(\bx)$ is an estimator for $\zeta_a(\bx) = E(Y_i^{(a)} \mid \bX_i = \bx) = E(Y_i \mid A_i = a, \bX_i = \bx)$.

\pagebreak

\label{lastpage}

\end{document}